\documentclass[preprint,authoryear,12pt]{elsarticle}
\usepackage{amsmath,amssymb}
\usepackage{graphicx}
\usepackage{bm}
\usepackage{multirow}
\usepackage{booktabs}
\usepackage{array}
\usepackage{tabularx}
\usepackage[utf8]{inputenc}
\usepackage[T1]{fontenc}
\usepackage{siunitx}
\usepackage[section]{placeins}
\usepackage{setspace}
\journal{}

\begin{document}

\begin{frontmatter}

\title{Computational Auditory Periphery Models: the Return of the Rodent}

\author[ugent]{Morgan Thienpont}\ead{morgan.thienpont@ugent.be}
\author[ugent]{F. Deloche}
\author[ugent]{S. Keshishzadeh}
\author[montpellier]{D. Kiselev}
\author[montpellier]{J. Bourien}
\author[montpellier]{J.-L. Puel}
\author[ohrc]{B. N. Buran}
\author[ncrar,ohsu]{N. Bramhall}
\author[ugent]{S. Verhulst}

\address[ugent]{Hearing Technology @ WAVES Team, Department of Information Technology, Ghent University, Ghent, Belgium}
\address[montpellier]{Institute for Neurosciences Montpellier, University of Montpellier, Montpellier, France}
\address[ohrc]{Oregon Hearing Research Center (OHRC), Department of Otolaryngology --- Head \& Neck Surgery, Oregon Health \& Science University, Portland, Oregon, USA}
\address[ncrar]{VA National Center for Rehabilitative Auditory Research (NCRAR), Veterans Affairs Portland Health Care System, Portland, OR, USA}
\address[ohsu]{Department of Otolaryngology/Head \& Neck Surgery, Oregon Health \& Science University, Portland, Oregon, USA}

\begin{abstract}
Animal experiments have provided many insights on auditory function, notably in cases of sensorineural hearing loss (SNHL). However, it is not always clear how these findings translate to the human auditory system, especially in clinically relevant contexts. Cross-species computational models of the auditory periphery can help bridge the gap between non-invasive human diagnostics and experimental evidence from animal studies. In this work we adapted a one-dimensional (1-D) nonlinear cochlear transmission-line (TL) model designed for the human auditory periphery \citep{verhulstcomputational2018} to mouse and gerbil, enabling a single computational framework for cross-species research on SNHL. Species-specific anatomical and physiological parameters --- including basilar membrane (BM) length and width, stapes area, middle-ear transfer functions, and characteristic-frequency range --- were adjusted to match each species' auditory periphery and hearing range. Other cochlear parameters were calibrated to reproduce realistic cochlear tuning and compressive growth. The adapted mouse and gerbil models were validated against experimental species-specific BM velocity level-growth characteristics, auditory-nerve (AN) tuning curves, and distortion-product otoacoustic emissions (DPOAEs). Simulated AN outputs reasonably matched empirical measurements, including realistic AN thresholds and frequency selectivity. However, the discrepancy between simulations and measurements became larger for cochlear sections closer to the base or apex. 

Simulations of auditory nerve synaptopathy reproduced observed differences in recorded auditory brainstem and envelope following responses from mice and gerbils with cochlear synaptopathy. However, OHC individualization of the mouse model based on DPOAEs failed to faithfully reproduce individual measured data, although inter-group differences in OHC damage were captured. Our findings demonstrate that biophysically grounded auditory periphery models can be translated across species while preserving realistic sound-coding properties and pathophysiological alterations. This approach refines the interpretation of animal data in specific hypotheses of human hearing, facilitates the development of new stimuli to test in rodents, and may enable in silico investigations of OHC loss, synaptopathy, and their functional consequences.
\end{abstract}

\end{frontmatter}

\section{Introduction}
Sensorineural hearing loss (SNHL) presents a significant challenge to auditory science and translational research, particularly in understanding how damage to cochlear structures affects auditory processing across species. Traditionally, studies have leveraged non-invasive measurements such as auditory brainstem responses (ABRs) and otoacoustic emissions (OAEs) to quantify the extent and nature of hearing loss in humans \citep{Kaushik2025OAE,Young2023ABR}. However, interpretation of these measurements is limited by the lack of direct confirmation of the underlying sources of dysfunction. Use of animal models, especially rodents such as mice and gerbils, offers the possibility to correlate non-invasive measurements with direct pathophysiological observations, thereby providing a more comprehensive understanding of auditory dysfunction \citep{Hickox2017TranslationalCochlearSynaptopathy}. Nevertheless, insights from animal models are not readily translatable to humans due to species-specific differences in cochlear processing.

Computational models that accurately capture inner-ear mechanics can simulate non-invasive measurements such as ABRs and OAEs, thereby revealing how specific cochlear pathologies shape these responses. Such models not only improve our understanding of normal auditory function, but also enable the systematic simulation of pathological conditions, providing a powerful tool for developing diagnostic and therapeutic strategies \citep{Buran2022SynapseCounts}. Individualized computational models provide a basis for developing personalized hearing-aid algorithms \citep{Wen2025dCoNNear}. Because hair cell and cochlear synapse loss cannot be quantified directly in living humans, existing computational models of the human auditory periphery have been used to link hypothesized cochlear damage patterns to non-invasive physiological measures  \citep{Buran2022SynapseCounts,keshishzadeh2021personalized}. This limitation is particularly critical for cochlear synaptopathy (CS): although synaptic loss is well established in animal models, there is still no consensus on its functional impact on human hearing. Active research on this topic aims to clarify the extent to which synaptopathy occurs in humans and its correlates with non-invasive markers \citep{Bharadwaj2019Synaptopathy,Bramhall2021ABR}. By enabling controlled manipulation of synaptic populations and quantifying their downstream effects on physiological and behavioural measures, computational models provide help to address these questions and design more sensitive, mechanistically grounded markers of synaptopathy. However, the absence of histological ground truth in humans limits the extent to which model-based predictions of cochlear damage can be independently verified, and thus constrains their broader acceptance \citep{Buran2022SynapseCounts,Bramhall2021ABR}. Adapting an existing human auditory periphery model \citep{verhulstcomputational2018} to two rodent species commonly used in auditory research (mouse and gerbil) allows us to refine the human model and validate its predictions against invasive animal data. Species-specific models tailored to laboratory animals also have the potential to reduce, and eventually replace, some animal experiments by enabling in silico exploration of cochlear pathologies.

The objective of this study was to translate a human auditory periphery model to mouse and gerbil so that the accuracy of the model could be validated by comparing simulated responses with measured responses from animals with known cochlear deficits. To meet this objective, a computational one-dimensional (1-D)  nonlinear cochlear transmission-line (TL) model, originally developed for human auditory research, was adapted to mouse and gerbil. This adaptation required modifications based on species-specific anatomical variations such as differences in stapes area, basilar membrane (BM) width and characteristic frequency (CF) range aligned with each animal’s tonotopic place-frequency map. In this study, we primarily concentrated on the mechanical aspects of the cochlea, without extensively exploring the dynamics of inner hair cells (IHCs) or cellular neural processes. Future research could build upon this foundation by investigating these components in greater detail.

The resulting mouse and gerbil models were evaluated by comparing simulated outputs of BM or auditory nerve (AN) responses and distortion-product otoacoustic emissions (DPOAEs) with experimentally measured physiological recordings from the respective species. Subsequently, the mouse model was utilized to simulate outer hair cell (OHC) loss and cochlear synaptopathy , characterized by the loss of cochlear synapses. To this end, the mouse model was individualized based on DPOAE recordings using a method previously applied to humans \citep{keshishzadeh2021dpoae} to estimate individual OHC damage. Furthermore, the simulated effect of ANF loss on auditory evoked potentials (AEPs) was compared with in-vivo mouse measurements and post-mortem synapse count data. For gerbils there was no OAE data available for OHC damage individualization. Therefore, AEPs were only simulated for one normal hearing (NH) and one synaptopathic profile.

\section{Methods and dataset}
The gerbil and mouse models were adapted from a 1-D human auditory periphery model described in a previous paper \citep{verhulstcomputational2018}. We first summarize the key features of the human model that are relevant to this work before explaining how the model was updated to reflect the anatomical and auditory response characteristics of mice and gerbils.

\subsection{Human model}
\subsubsection{Overview}
\begin{figure}
  \centering
  \includegraphics[scale=0.6]{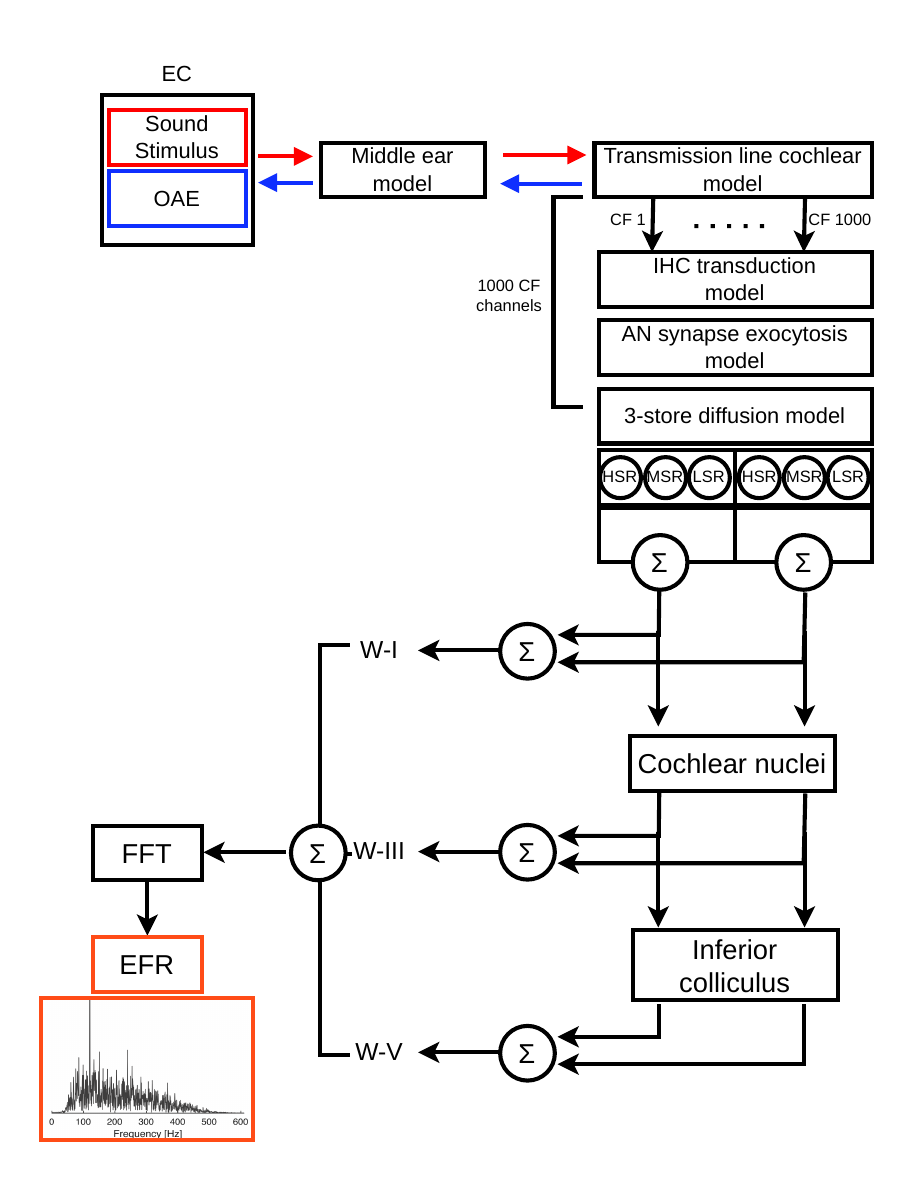}
  \caption{\textbf{Schematic overview of the computational auditory model } Diagram showing the main components of the human auditory periphery model \citep{verhulstcomputational2018} for simulating normal or hearing loss affected auditory responses at multiple stages along the auditory pathway with: EC = ear canal, OAE = otoacoustic emissions, IHC = inner hair cell, AN = auditory nerve, FFT = Fast Fourier Transform, EFR = envelope following response, CF = characteristic frequency, HSR = high spontaneous rate, MSR = medium spontaneous rate, LSR = low spontaneous rate fibers. W-I = ABR wave I, W-III = ABR wave III, W-V = ABR wave V. Red arrows indicate the forward traveling signal and blue lines the backward travelling signal. }
  \label{fig:model}.
\end{figure}

The human auditory periphery model (Figure \ref{fig:model}) first simulates how the acoustic stimulus passes through the middle-ear (ME), modeled as a first order Butterworth bandpass filter \citep{verhulstcomputational2018}. This is followed by a 1-D TL model which simulates the cochlear traveling wave along the BM. The BM is divided in cochlear sections, with each section corresponding to one CF. Irregularities in OHC gain are optionally added to simulate the reflections-source component of OAEs \citep{Verhulst2012}. The mechanically simulated BM velocity feeds into the inner hair cell (IHC) model which gives a biophysical description of the IHC membrane potential, thereby providing the mechanical-to-electrical
transduction. Further, the input to the AN synaptic complex is the IHC receptor potential $V_{\mathrm{IHC}}$, which controls the synaptic exocytosis rate. First the AN-synapse model computes the vesicle release and firing probability, and from that it produces the single-fiber firing rate \citep{altoe2018effect}. Auditory nerve fiber (ANF) responses are summed across different CFs that span the cochlear frequency map. These ANFs are divided into three subtypes: low-, medium-, and high-spontaneous-rate fibers (LSR, MSR, and HSR, respectively), with spontaneous firing rates (SRs) of 1, 10, and 68.5 spikes/s, respectively \citep{verhulstcomputational2018}. Note, the number of ANFs can be non-integer values due to continuous ANF distribution functions. In this case the ANF response is multiplied by the (non integer) number of ANFs and summed afterwards. This summation is passed to ABR generators that represent different auditory processing centers including the cochlear nucleus (CN) and inferior colliculus (IC). Envelope following responses (EFR) are then modeled by summing the weighted population responses from the AN, CN and IC \citep{carney2004phenomenological}. Weights are determined using the relative amplitudes of Wave I (W-I), Wave III (W-III), and Wave V (W-V) of the ABR. Hearing loss is simulated by reducing mechanical cochlear gain in the TL (corresponding to OHC loss) and by removing subtypes of ANFs (corresponding to CS).
 
\subsubsection{BM mechanics}\label{sec:BM mechanics}

The TL model is based on a time-domain implementation of Zweig's description of cochlear admittance \citep{zweigfinding1991}, which is grounded in the principle of local scaling symmetry. This time-domain model employs the TL formulation from \cite{altoetransmission2014}, incorporating cochlear tapering to achieve a resistive characteristic input impedance \citep{Verhulst2012,verhulstcomputational2018}.
 The TL impedances are given by: 
\begin{equation}
Z_{s_{n}}(s) = \omega_{n} M_{s_0} s
\label{eq:1}
\end{equation}
\begin{equation}
Y_{p_{n}}(s) = \frac{1}{Z_{p_{n}}(s)} = s \left[\omega_n M_{p_{0}} (s^2 + \delta_n n + 1 + \rho_n e^{-\mu_n s})\right]^{-1}
\label{eq:2}
\end{equation}
Subscripts \( s \) and \( p \) denote serial and shunt (parallel) parameters, respectively. In particular, $Z_{s_{n}}$ is the series impedance and $Y_{p_{n}}$ is the shunt admittance, also corresponding to the BM admittance. \( n \) is the cochlear section number, ranging from 1 to 1000. The constants $M_{s_0}$ and $M_{p_0}$ denote the longitudinal acoustic mass of the cochlear fluids and the phenomenological mass associated with the BM admittance, respectively (Table I in \cite{Verhulst2012}). Although the BM width and height are not defined specifically across CF, they are functionally incorporated in $Y_{p_n}$.

The variable \( s = i\omega/\omega_c \) is normalized by the cochlear section's CF. The parameters \( \delta \), \( \rho \), and \( \mu \) specify damping, stiffness, and delay, defined by the cochlear filter's primary double pole \( \alpha^* \) \citep{Shera20012}. The double pole \( \alpha^* \) determines the  sharpness of the cochlear filters. The expressions
for the pressure difference $p$ over one section of the cochlea $dx$ in function of the volume velocity $U$ is determined by the equations \citep{zweigfinding1991}: 

\begin{equation}
\frac{dp}{dx} = -Z_s U
\label{eq:zweig1}
\end{equation}

\begin{equation}
\frac{dU}{dx} = - \frac{p}{Z_p}
\label{eq:zweig2}
\end{equation}

A complete mathematical derivation is beyond the scope of this paper and can be found in earlier work \citep{verhulst2010phd,zweigfinding1991}. Following equations govern the TL model: cochlear section $n$ is coupled to sections $n-1$ and $n+1$ through:

\begin{equation}
\dot{U}_n = \frac{1}{dx M_{sn}}(p_{n-1} -2p_n +p_{n-1})
\label{eq:zweig3}
\end{equation}

The time derivative of the volume velocity in section $n$ is given by
\begin{equation}
\dot{U}_n = \ddot{y}_n \, dx,
\end{equation}
where $y_n$ denotes the displacement of the basilar membrane. This results in:

\begin{equation}
\ddot{y}_{n} = \frac{1}{b \space dx^2 M_{sn}}(p_{n-1} -2p_n +p_{n-1})
\label{eq:zweig3}
\end{equation}

Zweig's TL linear equations are  extended to a nonlinear version by dynamically shifting the double-pole \( \alpha^* \) horizontally in the \( s \)-plane \citep{Verhulst2012}. This shift broadens cochlear filters by moving the poles away from the imaginary axis in response to changes in stimulus intensity, while maintaining the BM velocity zero-crossings \citep{Shera2001, verhulstcomputational2018}. The position of \( \alpha^* \) determines the \( \delta \), \( \rho \), and \( \mu \) parameters in equation \ref{eq:2}, based on a derivation in \cite{Shera20012}. The corresponding equations are summarized below:
\setlength{\abovedisplayskip}{2pt}
\setlength{\belowdisplayskip}{2pt}

\noindent
\begin{minipage}{.5\textwidth}
\begin{equation}
    \delta = 2 (\alpha^* - a)
    \label{eq:3}
\end{equation}
\end{minipage}%
\begin{minipage}{.5\textwidth}
\begin{equation}
    \mu = \frac{1}{2\pi a}
    \label{eq:4}
\end{equation}
\end{minipage}

\noindent
\begin{minipage}{.5\textwidth}
\begin{equation}
    \rho = 2a \sqrt{1 - \left(\delta / 2\right)^2} \, e^{-\alpha^*/a}
    \label{eq:5}
\end{equation}
\end{minipage}%
\begin{minipage}{.5\textwidth}
\begin{equation}
    a = c^{-1} \left( \alpha^* + \sqrt{\alpha^{*2} + c (1 - \alpha^{*2})} \right)
    \label{eq:6}
\end{equation}
\end{minipage}
with constant $c = 120.9$. This $\alpha^*$-trajectory is determined based both on stimulus level and the CF, in order to model cochlear compressive response growth and the broadening of the cochlear filters and thus account for frequency tuning in humans \citep{Shera2010}. The key parameters for each CF in this process are:

\begin{enumerate}
    \item The active pole $\alpha^*_A$, representing the sharpest cochlear filters under low-level stimulation.
    \item The passive pole $\alpha^*_P$, corresponding to broader cochlear filters for high-level stimulation or post-mortem conditions.
    \item The intensity-dependent growth of basilar membrane velocity, $v_{\mathrm{BM}}$, characterized by a compression slope ($C$) of 0.31 dB/dB in the human model. 
    \item The compression threshold ($v_{\mathrm{BM,30dB}}$), which defines the level at which $v_{\mathrm{BM}}$ begins to grow compressively as the stimulus level increases.
\end{enumerate}

The compression threshold, $v_{\mathrm{BM,30dB}}$, is derived from the peak $v_{\mathrm{BM}}$ response to a 1-kHz pure tone at 30 dB SPL, measured at the cochlear section with the corresponding CF. The nonlinearity describing the $\alpha^*$-trajectory as a function of $v_{\mathrm{BM}}$ is modeled using a hyperbolic interpolation method \citep{Verhulst2012}. Specifically, as $v_{\mathrm{BM}}$ increases from $v_{\mathrm{BM,30dB}}$ to $v_{\mathrm{{BM,P}}}$, the pole $\alpha^*$ transitions between the active pole $\alpha^*_A$ and the passive pole $\alpha^*_P$ following:

\begin{equation}
\alpha^* = \alpha^*_A + \frac{x_p}{A} \sin \theta + \frac{y_p}{A} \cos \theta;
\end{equation}

Here, $\theta$ is defined as:

\begin{equation}
\theta = 0.5 \cdot \tan^{-1} 
\left( \frac{A \left( \alpha^*_P - \alpha^*_A \right)}{\frac{v_\mathrm{{BM,P}}}{v_{\mathrm{BM,30dB}}} - 1} \right)
\end{equation}

with a smoothing factor $A = 100$. The value of $v_{\mathrm{{BM,P}}}$ is computed from the fitting parameter $v_{\mathrm{BM,30dB}}$ to achieve the desired compression slope $C$ for $v_{\mathrm{BM}}$. $x_p$ normalizes and transforms the input $v_{\mathrm{BM}}$, while $y_p$ is defined by the conjugate hyperbola function:

\begin{equation}
\left( x_p / \alpha_p \right)^2 + \left( y_p / \beta_p \right)^2 = 1
\end{equation}

Furthermore, $x_p$ is expressed as:

\begin{equation}
x_p = \left( \frac{|v_\mathrm{{BM}}|}{v_{\mathrm{BM,30dB}}} - 1 \right) \frac{\cos \theta}{\cos 2\theta}.
\end{equation}

The parameters $\alpha_p$ and $\beta_p$ represent the slope of the hyperbola's asymptote, based on the focal point position ($F_p$), which is given by:

\begin{equation}
F_p = A \frac{\alpha^*_A}{v_\mathrm{{BM,P}} / v_\mathrm{{BM,30dB}}}
\end{equation}

Then, $\alpha_p$ and $\beta_p$ are computed as:

\begin{equation}
\alpha_p = F_p \cos \theta
\end{equation}

\begin{equation}
\beta_p = F_p \sin \theta
\end{equation}

It is important to note that the $\alpha^*$-trajectory increases with the input $v_{\mathrm{BM}}$, which results in progressively more damped values for $y_p$. The CF-dependence of alpha is modeled by a power law to match the desired cochlear frequency selectivity  defined by $Q_{\mathrm{ERB}}$. This becomes \citep{Shera2010}:

\begin{equation}
Q_\mathrm{{ERB,n}} = 11.46 \left( \mathrm{CF}_n / 1000 \right)^{0.25}
\label{eq:human_erb}
\end{equation}

\subsubsection{Simulating outer-hair-cell deficits}

Cochlear gain loss can be simulated by adjusting the parameter $\alpha_A$ in a CF-dependent manner. This allows to model hearing deficits arising from stereocilia damage, OHC loss, or metabolic decline (e.g., presbycusis). Since cochlear gain and $Q_{\mathrm{ERB}}$ are inherently linked in a 1-D TL model (equations 1 and 2), $\alpha_A$ can reflect a BM gain reduction corresponding to a specific hearing loss [dB HL]. As gain and bandwidth are tied to $\alpha_A$, BM filters in the hearing loss models also show reduced $Q_{\mathrm{ERB}}$. Hearing loss can be simulated by using $\alpha_A$ profiles corresponding to specific hearing losses (HL), while other BM nonlinearity parameters remain fixed. This is done by fitting $\alpha_A$ at cochlear section n with resonance frequency $f_n$, so that the reduction in gain with respect to the normal hearing pole profile equals the required HL at $f_n$. For example, if a HL of 20 dB at 1 kHz is needed, $\alpha_A$ at the cochlear section with CF = 1 kHz will be fitted so that the gain is reduced by 20 dB with respect to the gain in the NH profile. Numerically, this corresponds to an increase in $\alpha_A$, which enhances damping at low input levels, thereby reducing the available gain and leading to elevated thresholds.

\subsubsection{ME impedance}
To prevent the breaking of cochlear symmetry, which could lead to erroneous low-frequency
standing waves \citep{Shera1991}, the $R_{\mathrm{ME}}$ (middle - ear resistance) was matched to the cochlear impedance for low frequencies, (i.e., it was set equal to the low frequency cochlear impedance).  This implies \citep{verhulst2010phd}:

\begin{equation}
R_{ME} = \sqrt{\omega_{c0}^{2} M_{p0} M_{s0}}
\label{eq:human_erb}
\end{equation}

with 

\begin{equation}
\omega_{c0} = 2 \pi f_0
\label{eq:human_erb}
\end{equation}

where $f_0$ is the resonance frequency at the first cochlear section.

\subsubsection{Inner-hair-cell transduction}
The translation of BM movement to IHC transduction was simulated by the biophysical model developed by Altoé et al. \citep{altoe2018effect}. The model includes basolateral outward $K^{+}$ currents within the nonlinear IHC basolateral membrane \citep{verhulstcomputational2018}.

The $v_\mathrm{{BM}} [m/s]$ is transformed into $d_\mathrm{{IHCcilia}} [m] $ by multiplying $v_{\mathrm{BM}}$ with the $\kappa_{\mathrm{IHC}}$ ($d_\mathrm{{IHCcilia}}/v_\mathrm{{BM}}$) parameter. This parameter brings the IHC cilia deflection into a realistic biophysical operating range for the IHC transduction stage. Furthermore, it shifts the AN thresholds
to lower input levels when increased, and to higher input levels when decreased. This is essential when fitting the model to obtain realistic AN responses.

\subsubsection{Auditory-nerve synapse}
The AN synapses are represented using a two-stage modeling approach. The driven neurotransmitter exocytosis rate is first computed, after which a three-store diffusion model simulates the synaptic neurotransmitter kinetics \citep{verhulstcomputational2018}. Because these parameters were not changed when adapting the model to animals, the mathematical formulations are not discussed here. A comprehensive description is available in \cite{verhulstcomputational2018}.

\subsubsection{Auditory nerve fibers}
The output of AN synapse simulations are translated to ANF instantaneous firing rate. The original version of the human model contained 13 HSR, 3 MSR and 3 LSR fibers per cochlear section when simulating normal-hearing responses. CS can therefore be modeled by lowering the number of each subtype of fibers per CF. A constant HSR, MSR and LSR count across cochlear sections (as is the case in \cite{verhulstcomputational2018}) is not realistic. For gerbils and mice, literature provides descriptions of ANF distribution profiles where the number of ANFs varies with frequency (e.g., \cite{Bourien2014,Buran2022SynapseCounts}). 

\subsubsection{AN, CN and IC}
ANF activity is summed across cochlear sections to simulate AN responses. This summation yields the compound action potential (CAP). When combined with a dipole model, the CAP can be transformed into the generator potential underlying ABR W-I \citep{melcher1996generators}.  As can be seen in figure \ref{fig:model}, W-I (AN) is followed by the CN and the IC , corresponding to the ABR W-III and W-V components, respectively. The CN and IC responses are simulated using a functional spherical bushy cell model \citep{carney2004phenomenological}. Equations \ref{eq:CN} and \ref{eq:IC} represent the composite response of spherical bushy models in both excitatory and inhibitory neural dynamics, crucial for realistic simulations of AN responses and their propagation through higher auditory pathways. The CN and IC model equations are:
\begin{equation}
    r_{\text{CN}}(t) = A_{\text{CN}} \cdot \left[ \frac{t}{\tau^{2}_{\text{exc}}} \cdot e^{-\frac{t}{\tau_{\text{exc}}}} \ast r_{\text{AN}}(t) - S_{\text{CN,inh}} \cdot \frac{t}{\tau^{2}_{\text{inh}}} \cdot e^{-\frac{t}{\tau_{\text{inh}}}} \ast r_{\text{AN}}(t-D_{\text{CN}}) \right]
    \label{eq:CN}
\end{equation}
    
\begin{equation}
    r_{\text{IC}}(t) = A_{\text{IC}} \cdot \left[ \frac{t}{\tau^{2}_{\text{exc}}}\cdot e^{-\frac{t}{\tau_{\text{exc}}}} \ast r_{\text{CN}}(t) - S_{\text{IC,inh}} \cdot \frac{t}{\tau^{2}_{\text{inh}}} \cdot e^{-\frac{t}{\tau_{\text{inh}}}} \ast r_{\text{CN}}(t-D_{\text{IC}}) \right]
    \label{eq:IC}
\end{equation}

The model uses response functions $r(t)$, amplitude scaling factors $A$, and excitatory and inhibitory time constants $\tau_{\text{exc}}$ and $\tau_{\text{inh}}$. The parameter $S_{\text{inh}}$ represents the relative CN/IC inhibition–excitation strength, while $r_{\text{AN}}(t)$ denotes the AN response function. Finally, $D$ specifies the CN/IC inhibition delay. To simulate  W-I, W-III and W-V responses within a realistic range, $r_\mathrm{{AN}}$, $r_\mathrm{{CN}}$, and $r_\mathrm{{IC}}$ are multiplied by the scaling factors $A_\mathrm{{W-I}}$, $A_\mathrm{{W-III}}$, and $A_\mathrm{{W-V}}$, respectively. These scaling factors are derived from empirical ABR measurements. Finally, the contribution of W-I, W-III, and W-V are summed together to yield the EFR.

\subsection{Translating the model framework to different species}

Translating the Verhulst et al. 2018 TL model to gerbils and mice requires several steps. First, species-specific anatomical and physiological data is needed to update the ME and cochlear model parameters. Further, AN tuning curves are required to determine the TL poles, as well as BM velocity measurements to fit the compression slope and knee point of the nonlinearities. Next, the ANF distribution of the animals needs to be estimated based on ANF measurements. Lastly, the ABR wave weights are determined based on click ABR recordings. When the models are calibrated, they can be validated by comparing them with measurements of AEPs of selected stimuli. 

\subsubsection{Sources of data}
\begin{figure}
  \centering
  \includegraphics[scale=0.25]{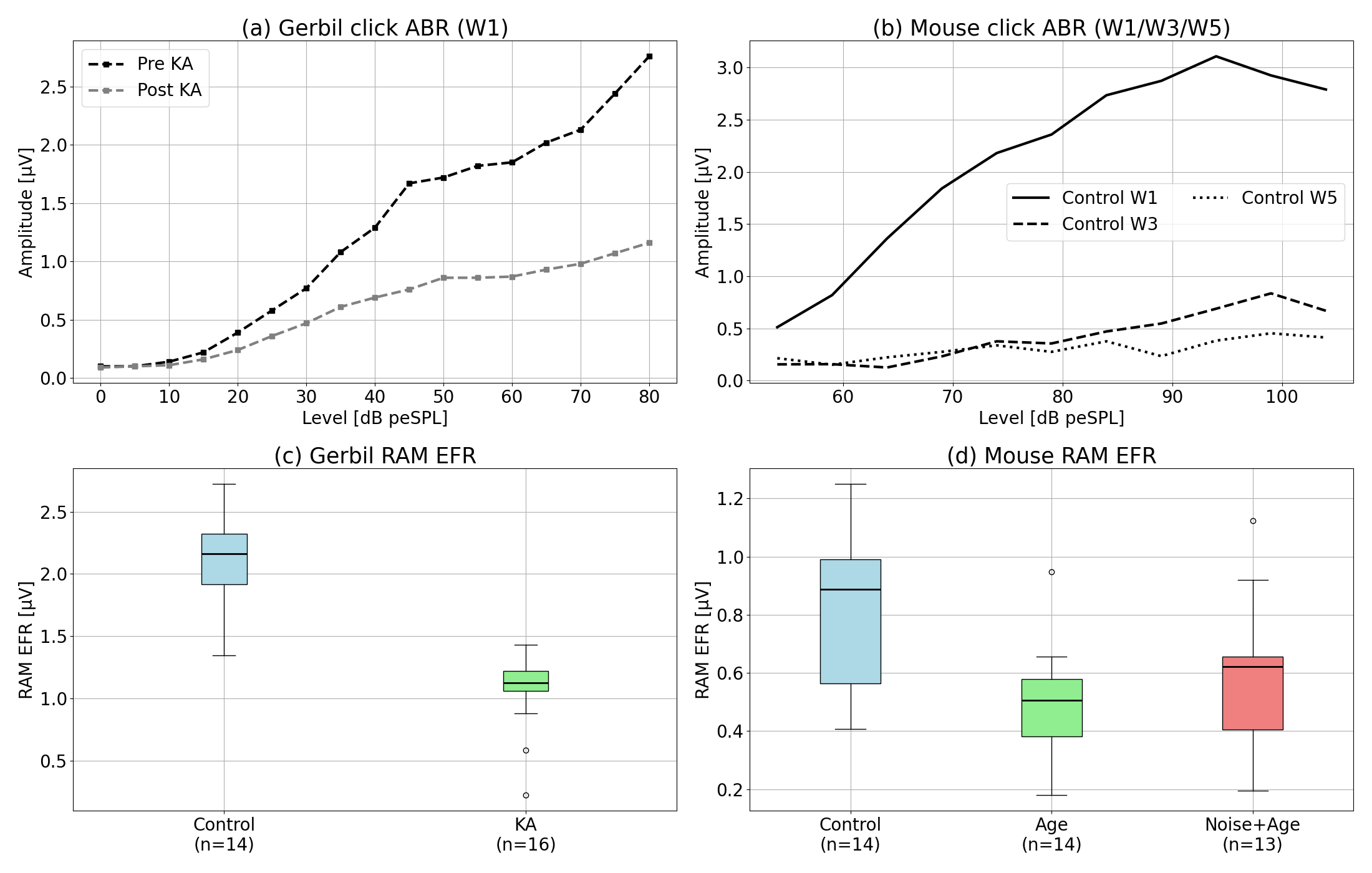}
  \caption{ \textbf{Animal dataset ABRs and EFRs} a) Experimental click ABRs (W1) of the gerbil reference dataset pre and post kainic acid (KA) administration. b) Experimental click ABRs (W1, W3, and W5) of the reference control mouse dataset. (c) Reference envelope-following responses (EFRs) to a rectangular amplitude-modulated (RAM) pure tone stimulus for the mouse and gerbil populations. (c) RAM EFRs measured in gerbils using a carrier frequency of 4~kHz and a modulation frequency of 116~Hz presented at an intensity of 70~dB SPL. Two experimental groups were tested: control and KA treated animals. (d) RAM EFRs measured in mice using a carrier frequency of 8~kHz and a modulation frequency of 110~Hz presented at an intensity of 70~dB SPL. Results are shown for three groups: young control, age, and age + noise exposed.}.
 \label{fig:Mouse_EFR_boxplot}.
\end{figure}

We combined anatomical information from literature with new physiological and histological measurements to adapt the human model to mouse and gerbil. Most anatomical parameters were taken from existing studies (Table~\ref{tab:table1}). When direct parameter values were unavailable, published datasets were used to fit the equations required by the model (e.g., the frequency dependence of $Q_{\mathrm{ERB}}$). Histological data (synapse ribbon counts) and physiological data (DPOAEs, click-evoked ABRs, and EFRs) were collected from 41 mice, with a subset of 37 animals previously reported in \citep{Bradmouse2025}. For gerbils, click-evoked ABR Wave~1 (W1) responses were measured in 36 animals before and two weeks after kainic acid (KA) administration, and EFRs were recorded in 19 control and 20 KA-treated gerbils \citep{Daniil}. EFRs in both species were recorded to a rectangular amplitude-modulated stimulus (RAM-EFR). Figures~\ref{fig:Mouse_EFR_boxplot}a and b show experimental click-evoked ABRs for gerbils and mice (W1 for gerbil, W1, W3, and W5 for mouse), and Figures~\ref{fig:Mouse_EFR_boxplot}c and d show distributions of RAM-EFR magnitudes for both species. The RAM stimulus definition and RAM-EFR metric are detailed in Section~\ref{sec:ram_efr_methods}.

\paragraph{Gerbil data}
Click-evoked ABR W1 peak-to-peak amplitudes were measured in control and KA-treated gerbils before and two weeks after KA administration (Figure~\ref{fig:Mouse_EFR_boxplot}a). RAM-EFR magnitudes were compared between the two groups (Figure~\ref{fig:Mouse_EFR_boxplot}c). KA induces synaptic loss and is used as a CS model in gerbil and budgerigar studies \citep{Diuba2025,Wilson2021EFRBudgerigar}. Gerbil data were collected at the Institute of Neurosciences of Montpellier.

\begin{figure}
  \centering
  \includegraphics[scale=0.35]{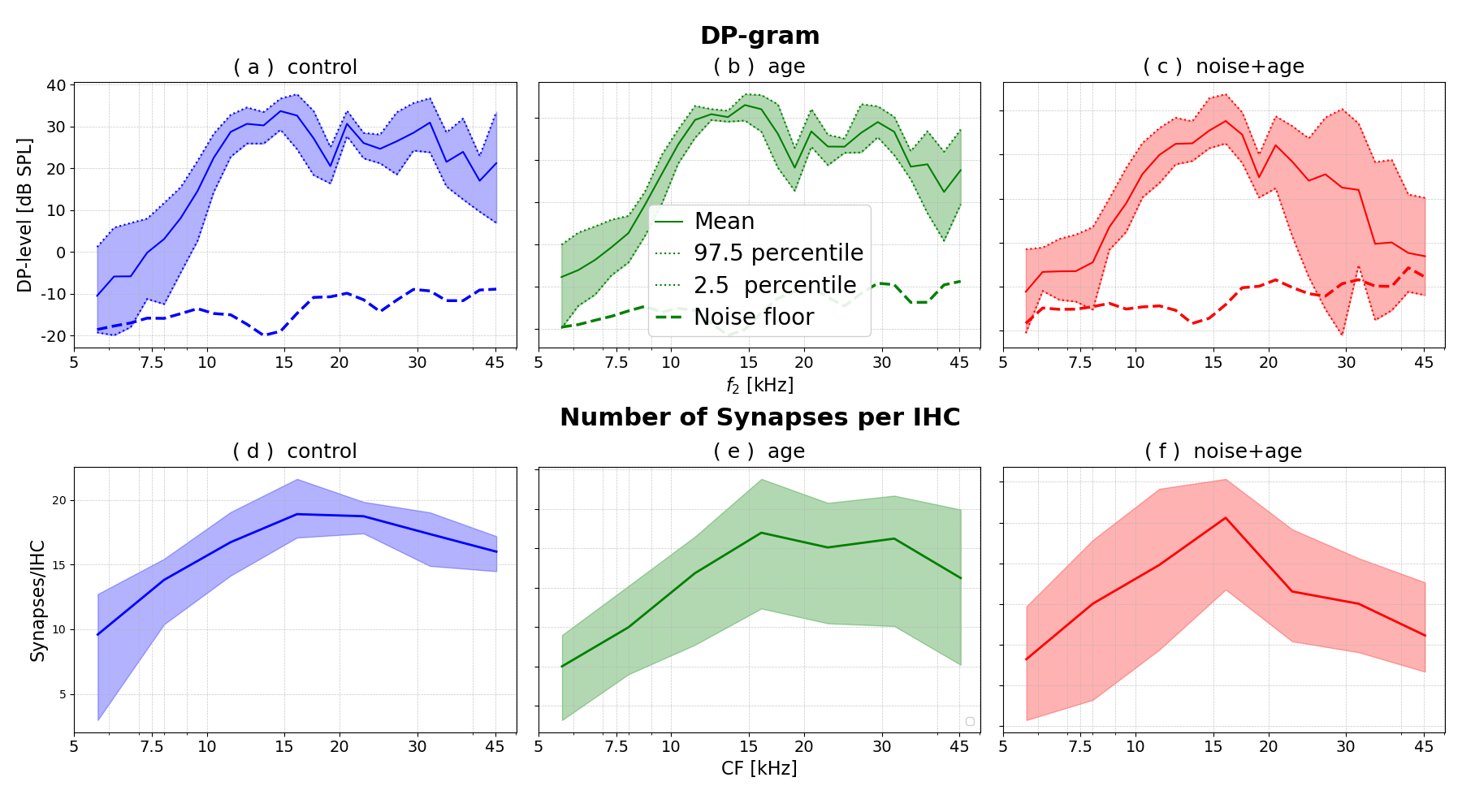}
  \caption{ \textbf{DPOAEs and synapse count of the mouse dataset} Animal data showing the average and 2.5 and 97.5 percentile intervals of distortion-product otoacoustic emission (DPOAE) levels of three mice groups, a) young control, b) age, and c) age + noise exposed, for an intensity level of 55 dB SPL for $f_1$ and 65 dB SPL for $f_2$. The ratio of $f_2$ to $f_1$ is 1.2. The DP-level is defined as the sound pressure level (dB SPL) measured at the frequency $2f_1 - f_2$. d-f) Synapses per IHC (inner hair cell) as a function of CF (characteristic frequency) for the three groups of mice. }
  \label{fig:ribbons_dpoae_groups}.
\end{figure}

\paragraph{Mouse data}
     The mouse dataset included control, aged, and aged plus noise-exposed groups. Figures~\ref{fig:ribbons_dpoae_groups}a and c summarize recorded DPOAEs used to individualize the BM pole profiles. DPOAEs were measured using stimulus levels of 55 dB SPL for $f_1$ and 65 dB SPL for $f_2$, with a fixed frequency ratio of $f_2/f_1 = 1.2$. The DP-level is defined as the sound pressure level (dB SPL) measured at the frequency $2f_1 - f_2$. Synapse ribbon counts per IHC are shown in Figures~\ref{fig:ribbons_dpoae_groups}d--f. Figure~\ref{fig:Mouse_EFR_boxplot}d presents RAM-EFR magnitudes across these groups, and Figure~\ref{fig:Mouse_EFR_boxplot}b shows click-evoked ABRs (peak-to-peak W1, W3, and W5). Mice were divided into three groups: young, noise-exposed, and aged. The young group comprised animals aged 10 to 42 weeks. Mice in the noise-exposed group were subjected to 101 dB SPL at 16 weeks of age and subsequently tested between 80 and 104 weeks. The aged group consisted of animals tested between 82 and 103 weeks of age. These data were provided by \cite{Bradmouse2025} and are available on Zenodo (\emph{https://zenodo.org/records/19455491}).

\subsubsection{Anatomical and physiological parameters}

\begin{table}[htbp]
\centering
\footnotesize
\setlength{\tabcolsep}{4pt}
\caption{\textbf{Adapted auditory model parameters} Anatomical and functional parameters of the human, gerbil and mouse model with: BM = basilar membrane, fs = sampling frequency, $d_{\mathrm{IHCcilia}}$ = inner hair cell stereocilia displacement, $v_\mathrm{{BM}} $ = basilar membrane velocity. Human parameters are as described in \cite{verhulstcomputational2018}.}
\label{tab:table1}
\begin{tabularx}{\textwidth}{@{}>{\raggedright\arraybackslash}p{0.28\textwidth}>{\raggedright\arraybackslash}p{0.21\textwidth}>{\raggedright\arraybackslash}p{0.24\textwidth}>{\raggedright\arraybackslash}p{0.21\textwidth}@{}}
\toprule
\textbf{Parameter} & \textbf{Human \citep{verhulstcomputational2018}} & \textbf{Gerbil} & \textbf{Mouse} \\
\midrule
BM width & 1 mm & 192 $\mu$m \citep{edgemorphology1998} & 125 $\mu$m \citep{santidevelopment2008} \\
BM height & 1 mm & 316 $\mu$m \citep{Yoon2009} & 170 $\mu$m \citep{keppelermultiscale2021} \\
Cochlear length & 35 mm & 12.1 mm \citep{Greenwood1990} & 6 mm \citep{Montgomery2016} \\
Helicotrema width & 1 mm & 0.5 mm \citep{Xia2018} & 0.15 mm \citep{MOUNTAIN2003} \\
Greenwood $\alpha$ & 61.675 & 189.65 \citep{muller1996cochlear} & 358.97 \citep{Greenwood1990} \\
Greenwood B ($f_B$) & 140.6 & 251.14 \citep{muller1996cochlear} & 816 \citep{Greenwood1990} \\
Greenwood A ($f_A$) & 20682 & 63079 \citep{muller1996cochlear} & 120849.7 \citep{Greenwood1990} \\
Stapes area & 3 mm$^2$ & 0.4 mm$^2$ \citep{Mason2016} & 0.143 mm$^2$ \citep{motallebzadehmouse2021} \\
Middle-ear gain & 18 dB & 25 dB \citep{Dong2009} & 37 dB \citep{motallebzadehmouse2021} \\
Middle-ear filter (cut-off) & 0.6--4 kHz & 1--20 kHz \citep{Ravicz2008} & 8--25 kHz \citep{motallebzadehmouse2021} \\
Middle-ear filter (order) & $2 \times 1$ & $2 \times 2$ & $2 \times 3$ \\
Sampling rate (fs) & 100 kHz & 200 kHz & 300 kHz \\
Compression slope & 0.4 & 0.6 & 0.4 \\
Knee point & 20 dB & 35 dB & 28 dB \\
$Q_{\mathrm{ERB}}$ ($f_0 = 1$ kHz) & $11.46 \cdot (f / f_0)^{0.25}$ & $3.26 \cdot (f / f_0)^{0.53}$ \citep{heeringa2024single} & $7.47 \cdot (f / f_0)^{0.18}$ \citep{Taberner2005} \\
$\kappa_{\mathrm{IHC}}$ $(d_{\mathrm{IHCcilia}}/v_\mathrm{{BM}})$ & 0.118 s & 7 s & 2 s \\
$A_\mathrm{{W\text{-}I}}$ & $6.28 \times 10^{-14}$ & $1.57 \times 10^{-13}$ & $1.33 \times 10^{-11}$ \\
$A_\mathrm{{W\text{-}III}}$ & $7.22 \times 10^{-14}$ & $2.23 \times 10^{-13}$ & $4.80 \times 10^{-12}$ \\
$A_\mathrm{{W\text{-}V}}$ & $3.52 \times 10^{-20}$ & $3.91 \times 10^{-13}$ & $4.08 \times 10^{-12}$ \\
\bottomrule
\end{tabularx}
\end{table}
Table \ref{tab:table1} shows an overview of the anatomical and functional parameters from the human model that were updated to generate the animal models, with the first column corresponding to the original human model. Gerbil and mouse anatomical features were obtained from literature. Because mice and gerbils hear substantially higher frequencies than humans, the simulation sampling rate was increased by a factor of two for the gerbil model and a factor of three for the mouse model to satisfy the Nyquist--Shannon criterion and avoid aliasing across the full audible range. The density of the cochlear fluid ($\rho_{fluid}$), the number of cycles of the traveling waves that were traversed before its maximum was reached (N) were left unchanged as these are fundamentals of the cochlear scaling symmetry principle across species \citep{Shera1991}. The place–frequency mapping of the cochlea is described by the Greenwood function \citep{Greenwood1990}:
\begin{equation}
f_n = f_A \, 10^{-\alpha n} - f_B
\label{eq:greenwood}
\end{equation}
The parameters $f_A$, $f_B$, and $\alpha$ were selected such that the resulting mapping spans the hearing range of the respective animal model.

\begin{figure}
  \centering
  \includegraphics[scale=0.18]{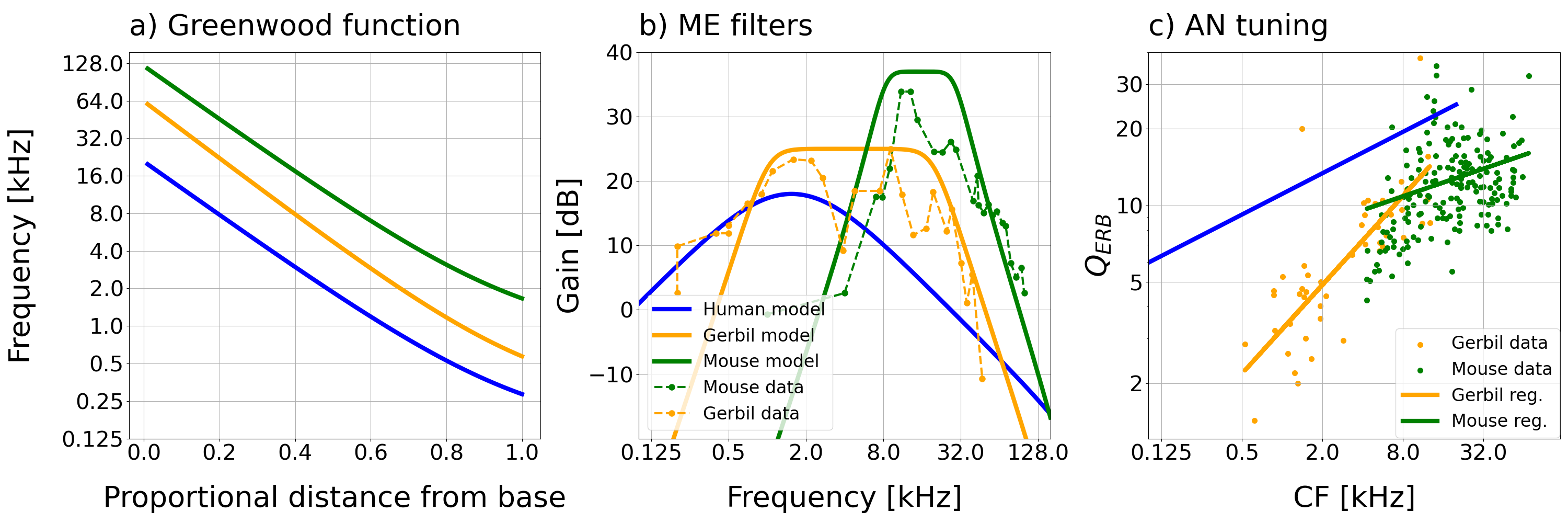}
  \caption{\textbf{Functional changes in auditory model hearing range} a) Relationship between tonotopic frequency and fractional distance from the base of the cochlea for each species as defined by the species-specific Greenwood function b) Gain profile of the middle - ear filter for each species compared to data from literature, i.e. rescaled stapes velocity relative to ear canal pressure for gerbil and forward cochlear pressure relative to umbo presssure for mice \citep{Ravicz2008,motallebzadehmouse2021}. c)  Regression models (reg.) describing the sharpness of tuning ($Q_{\mathrm{ERB}}$) as a function of the characteristic frequency (CF). These regression models are used to set the animal model poles by approximating species-specific behavior. The regressions are shown alongside measurements from single-unit auditory nerve responses (individual markers) for gerbils \citep{Schmiedt1989} and mice \citep{Taberner2005}. The relationship $Q_{\mathrm{ERB}} = 2\,Q_{\mathrm{10dB}}$ is assumed. Parameters are reported in Table \ref{tab:table1}.}
  \label{fig:panelA}.
\end{figure}

Figure~\ref{fig:panelA} shows the resulting Greenwood functions (panel~a), ME filters (panel~b), and $Q_{\mathrm{ERB}}$ (panel~c). The power law describing sharpness of tuning ($Q_{\mathrm{ERB}}$) was derived by fitting a regression line to $Q_{\mathrm{ERB}}$ estimates derived from single-unit auditory nerve data from gerbil and mouse \citep{Schmiedt1989,Taberner2005}. The regression line was fitted on a log--log scale, such that 
$\log(Q_{\mathrm{ERB}}) = \log\!\left(\beta_0 (f / f_0)^{\beta_1}\right)$. 
Since sharpness of tuning for auditory nerve data is typically reported as the width of the tuning curve at 10 dB relative to threshold at characteristic frequency ($Q_\mathrm{{10dB}}$), the values were converted to $Q_{\mathrm{ERB}}$ by multiplying $Q_\mathrm{{10dB}}$ by a factor of 2. This relationship results from the frequency analysis of impulse responses simulated with
our model. This differs from the theoretical $Q_{\mathrm{ERB}}$ to $Q_\mathrm{{10dB}}$ ratio of 1.78 that is found when considering gammatone filters \citep{Thienpont2024Zenodo, Shera2002PNAS}. The models were fit to the $Q_{\mathrm{ERB}}$ functions by adjusting the parameters in equations \ref{eq:3}, \ref{eq:4}, \ref{eq:5}, \ref{eq:6}. Final $Q_{\mathrm{ERB}} (CF)$ functions can be found in Table~\ref{tab:table1}.

\subsubsection{BM-response poles}

Because the admittance poles determine BM tuning (Section \ref{sec:BM mechanics}), they were re-estimated for each species to reflect the desired $Q_{\mathrm{ERB}}$ (CF) profiles. The same general approach as in the human model was applied, but using the species-specific $Q_{\mathrm{ERB}}$ targets: BM velocity at a mid-cochlear section was simulated for a 0\, peSPL click across a physiologically realistic and mathematically stable pole range. The click responses were analyzed to establish the relationship between the pole values and the effective $Q_{\mathrm{ERB}}$. This relationship was subsequently used to approximate the power-law dependence of $Q_{\mathrm{ERB}}$ on CF, as described in the previous subsection, enabling determination of the model poles within the low-level linear regime.

\subsubsection{BM compressive nonlinearities}
The next step was to adjust the BM nonlinearities for each species. Mechanical data from optical coherence tomography (OCT) \citep{He2022, Dewey2019} was used to set the knee points and compressive slopes of the desired BM velocity input/output function. Frequency selectivity in the gerbil and mouse model were significantly lower than for the human model (Figure ~\ref{fig:panelA} c). These characteristics caused several shortcomings that had to be addressed. Lower frequency selectivity ($Q_{\mathrm{ERB}}$) implies a higher starting pole, resulting in a limited range for  compression.  Given the inherent coupling between tuning and gain in 1-D cochlear transmission line models, the value of the starting pole (determined by $Q_{\mathrm{ERB}}$) determines the maximum compression that can be generated at that same CF. For species with poor frequency selectivity, this means that a regular 1-D model structure is unable to capture the experimentally observed compression at high levels.

\subsubsection{Inner-hair-cell and auditory nerve transduction}
The $\kappa_{\mathrm{IHC}}$ parameters for the gerbil and mouse models were primarily based on the dynamic range of the ANF responses and their resulting influence on AN tuning curves. The IHC model was kept invariant across CF. Similarly, the HSR, MSR, and LSR ANFs were each represented by a single parameter set that remained constant across CF and between models.

\subsubsection{Auditory nerve fiber subtype profiles}
\begin{figure}
  \centering
  \includegraphics[scale=0.3]{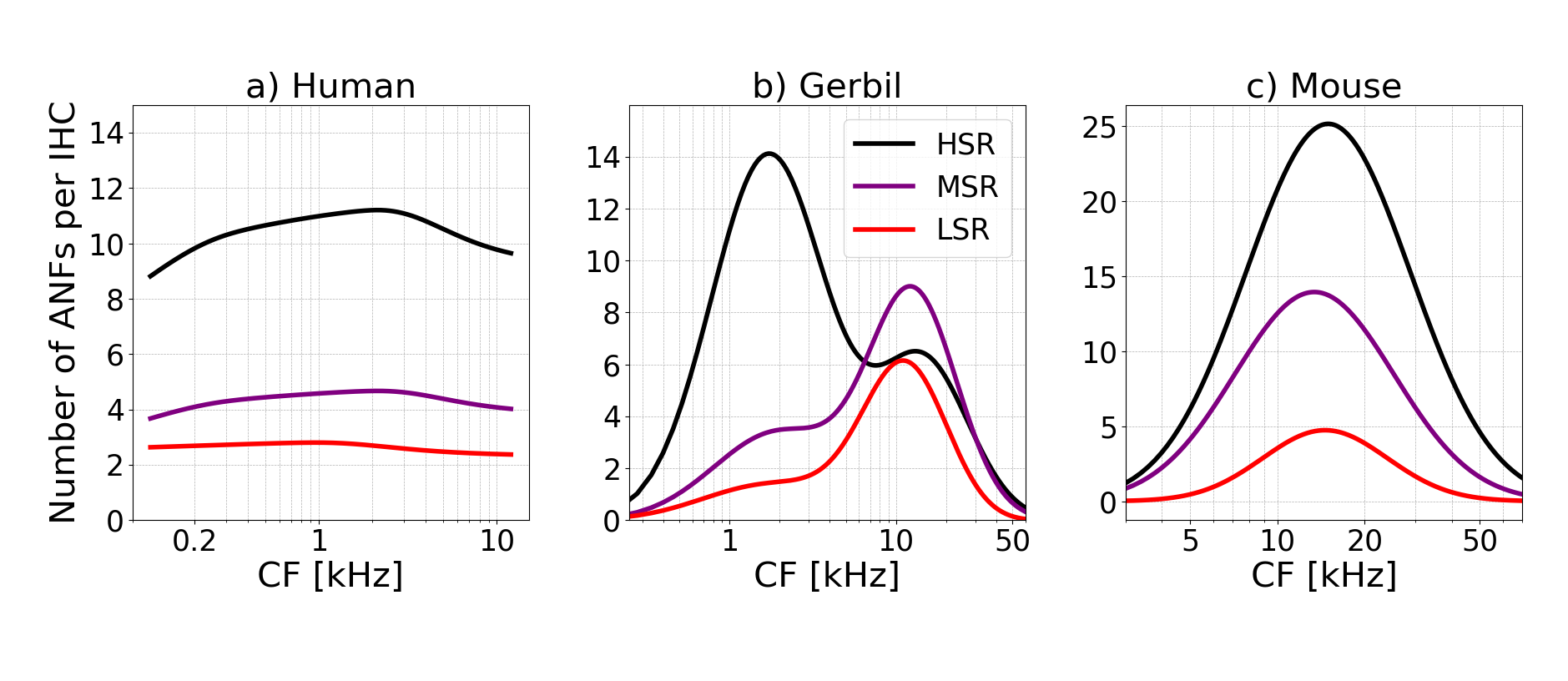}
  \caption{ \textbf{Auditory nerve fiber distributions} Number of auditory nerve fibres (ANFs) of the high- (HSR), medium- (MSR), and low-spontaneous-rate (LSR) types per inner hair cell (IHC) as a function of characteristic frequency (CF) in the three normal-hearing models: (a) human, (b) gerbil, and (c) mouse.}.
  \label{fig:ANF_dist}.
\end{figure}

Different species have different ANF distributions or profiles along CFs of the cochlea. This is reflected in the HSR, MSR and LSR count per cochlear section in the model. In the current modeling framework, HSR, MSR, and LSR fibers are assigned spontaneous firing rates of 68.5, 10, and 1 spikes/s, respectively. Their corresponding peak firing rates are 3000, 1000, and 800 spikes/s. ANF profiles for each model are shown in Figure \ref{fig:ANF_dist}. The human profile and its estimation has been described in a previous paper \citep{keshishzadeh2021personalized}. 
Gerbils have a distinct ANF distribution profile that has two peaks in the number of fibers across CF. Therefore, the HSR, MSR, and LSR distributions are approximated by fitting a double Gaussian distribution to data provided in  \cite{Bourien2014}. In the mouse model, a regular Gaussian distribution is applied to match experimental observations on the reference dataset. In both cases the distributions are defined with respect to the logarithm of CF.

\begin{table}[htbp]
\centering
\footnotesize
\setlength{\tabcolsep}{4pt}
\caption{Parameters describing the distribution of auditory nerve fiber (ANF) types—high (HSR), medium (MSR), and low (LSR) spontaneous rate fibers in humans, gerbils, and mice. Gerbil distributions use cumulative double-Gaussian models (HSR, HSR+MSR, HSR+MSR+LSR) based on \cite{Bourien2014}. Mouse distributions use single-Gaussian models with parameters expressed in natural-log frequency space, approximation based on reference dataset.}
\label{tab:anf_parameters}
\begin{tabularx}{\textwidth}{@{}>{\raggedright\arraybackslash}p{0.12\textwidth}>{\raggedright\arraybackslash}p{0.18\textwidth}X@{}}
\toprule
\textbf{Species} & \textbf{Fiber type} & \textbf{Model and parameters} \\
\midrule
Gerbil & Model & Double Gaussian (cumulative representation) \\
Gerbil & HSR & $\theta_{\mathrm{HSR}} = [0.1411,\ 0.5461,\ 0.7942,\ 0.0606,\ 2.6848,\ 0.6244]$ \\
Gerbil & HSR + MSR & $\theta_{\mathrm{HSR+MSR}} = [0.1747,\ 0.5671,\ 0.8103,\ 0.1465,\ 2.5948,\ 0.6206]$ \\
Gerbil & HSR + MSR + LSR & $\theta_{\mathrm{HSR+MSR+LSR}} = [0.1887,\ 0.5695,\ 0.8191,\ 0.2044,\ 2.5417,\ 0.6134]$ \\
\addlinespace
Mouse & Model & Single Gaussian \\
Mouse & HSR & $\theta_{\mathrm{HSR}} = [0.2516,\ 9.6127,\ 0.6522]$ \\
Mouse & MSR & $\theta_{\mathrm{MSR}} = [0.1394,\ 9.4959,\ 0.6325]$ \\
Mouse & LSR & $\theta_{\mathrm{LSR}} = [0.0473,\ 9.5971,\ 0.4952]$ \\
\bottomrule
\end{tabularx}
\end{table}

The double Gaussian function used in the gerbil ANF profile consists of
the sum of two single-Gaussian components:
\begin{equation}
G(x; \theta_{11},\theta_{12},\theta_{13},\theta_{21},\theta_{22},\theta_{23})
= g(x; \theta_{11},\theta_{12},\theta_{13})
+ g(x; \theta_{21},\theta_{22},\theta_{23}),
\label{eq:double_gaussian_ln}
\end{equation}

with:
\begin{equation}
g(x; \theta_{1},\theta_{2},\theta_{3})
= \theta_{1} \exp\!\left( -\frac{(x - \theta_{2})^2}{2\theta_{3}^2} \right),
\label{eq:single_gaussian_ln}
\end{equation}
where $\theta_{1}$ is the amplitude, $\theta_{2}$ the center position, and $\theta_{3}$ the spread
parameter. 

The gerbil ANF distribution, $F_{\mathrm{gerbil}}(f)$, is then described as the natural logarithm of frequency (in kHz):
\begin{equation}
F_{\mathrm{}}(f) = G\!\left( \ln(f) \right).
\label{eq:gerbil_ln}
\end{equation}

Separate distributions for gerbil HSR, MSR, and LSR fibers are obtained by
taking differences between the cumulative double-Gaussian models:
\begin{align}
\mathrm{HSR}(f) &= G_1(\ln f), \\
\mathrm{MSR}(f) &= G_2(\ln f) - G_1(\ln f), \\
\mathrm{LSR}(f) &= G_3(\ln f) - G_2(\ln f),
\end{align}
yielding three independent ANF populations. This method was applied in order to match the data from  \cite{Bourien2014} where a cumulative distribution is described.

\subsubsection{AN, CN and IC}
The contributions of the AN, CN, and IC to the summed EFR are weighted by the $A_\mathrm{{W\text{-}I}}$, $A_\mathrm{{W\text{-}III}}$, and $A_\mathrm{{W\text{-}V}}$ scaling factors, respectively. For the gerbil model, these scaling factors were calibrated to match the click-evoked ABRs reported in \cite{burkard1989stimulus} at 80~dB peSPL. The same procedure was applied to the mouse model, where the scaling factors were adjusted to the reference mouse ABR dataset. For W1 and W3, calibration was performed using responses at 89~dB peSPL, whereas the W5 weight was calibrated at 79~dB peSPL, which yielded a better overall fit than using 89~dB peSPL.

\subsubsection{Mouse model individualization based on DPOAEs}
For the mouse model, individualization was performed in two stages: first by estimating OHC damage, and the corresponding starting pole values across CF, from DPOAEs, and then by adjusting ANF distributions based on synapse counts (\ref{sec:ANF_ind}).
The TL model can be individualized based on audiogram, DPOAE thresholds (DPTH), or DP-levels \citep{keshishzadeh2021dpoae}. We opted here to individualize the mouse model based on measured DPOAEs (e.g. DP-levels). The same method that has been elaborated in \cite{keshishzadeh2021dpoae}, is applied here to the mouse model. It uses a deep neural network (DNN), trained on simulated DPOAEs, to map measured DPOAEs to poles in the model. The $f_2$ of measured DPOAEs, used to individualize the model, covered frequencies between 5 and 46 kHz. The levels of $f_1$ and $f_2$ were 55 and 65 dB SPL, respectively. We only applied  OHC individualization to the mouse model as the reference mouse data set contained DPOAEs, noise-exposed subjects, and age groups whereas this was not the case for the gerbil reference data we had available for this study. 

\subsubsection{Mouse model individualization based on synapse counts}
\label{sec:ANF_ind}
The mouse model was individualized for each animal by scaling the ANF distribution according to the ratio of the animal’s ribbon count per IHC at CF = 8 kHz to the maximum ribbon count per IHC at the same CF observed in this population. We chose 8 kHz as the anchor CF because the RAM EFR in the measured and simulated mouse data has an $f_c$ of 8 kHz as well:

\begin{equation}
\mathrm{ANF}_i(8\,\mathrm{kHz}) = \frac{R_i(8\,\mathrm{kHz})}{\max\!\left(R(8\,\mathrm{kHz})\right)} \cdot \mathrm{ANF}_{\mathrm{NH}}(8\,\mathrm{kHz})
\label{eq:ANFi}
\end{equation}
with 
\begin{equation}
R_i(f) = \frac{\text{number of ribbons}}{\text{IHC}}(f)
\label{eq:Ri}
\end{equation}

where $ANF_{i}$ is the ANF distribution of animal $i$ and $ANF_{NH}$ is the NH ANF distribution shown in figure \ref{fig:ANF_dist} c.

Population responses, such as AEPs, can be linked to individual ANF profiles and can serve as a marker for synaptopathy. However, non-invase AEP measurements in humans lack the IHC ribbon count as ground truth needed to verify the marker and its quantitative relation to the individual ANF survival. Currently, synaptopathy in humans can be derived from electrophysiological markers such as the EFR to a RAM stimulus in combination with iterative model simulations of a range of possible CS profiles \citep{keshishzadeh2021personalized}. The RAM EFR has been shown to function as a marker for synaptopathy in animal models such as the budgerigar \citep{Wilson2021EFRBudgerigar} and the gerbil and mouse reference data in this paper.

\subsubsection{RAM stimulus and EFR computation}
\label{sec:ram_efr_methods}

\begin{figure}
  \centering
  \includegraphics[scale=0.3]{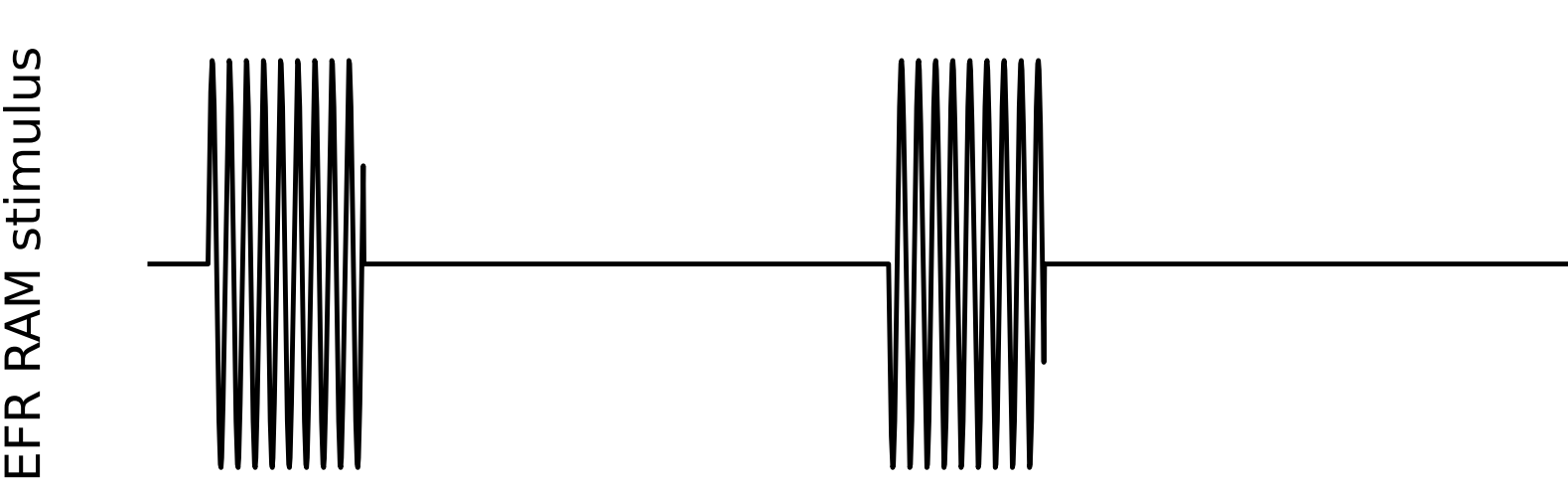}
  \caption{ \textbf{RAM EFR stimulus} Rectangular-amplitude-modulated pure tone (RAM) stimulus. This stimulus is used to measure the envelope-following response (EFR).}
  \label{fig:RAMEFR}
\end{figure}

A rectangular amplitude-modulated (RAM) stimulus consists of a carrier frequency $f_c$ and a modulation frequency $f_m$. Figure \ref{fig:RAMEFR} shows the general shape of a RAM stimulus. The measured and simulated RAM-EFRs were computed as described in \cite{keshishzadeh2021personalized} by summing the noise-floor–corrected spectral magnitudes $M_{f_k}$ at $f_m$ and its first three harmonics, $f_2$ to $f_4$. Because a large proportion of the fifth-harmonic components in the measurements fell below the noise floor, the summation in (\ref{eq:RAM_EFR}) was restricted to the fourth harmonic of $f_m$.

\begin{equation}
\mathrm{RAM}_{\mathrm{EFR}} = \sum_{k=1}^4 M_{f_k} \space, \space f_k = f_m \cdot k
\label{eq:RAM_EFR}
\end{equation}

The modulation depth and duty cycle were 100\%, and 25\%, respectively. Note that for simulations, there is no noise-floor correction. 

\section{Model Validation}
\subsection{BM response}
\begin{figure}[h]
  \centering
  \includegraphics[scale=0.3]{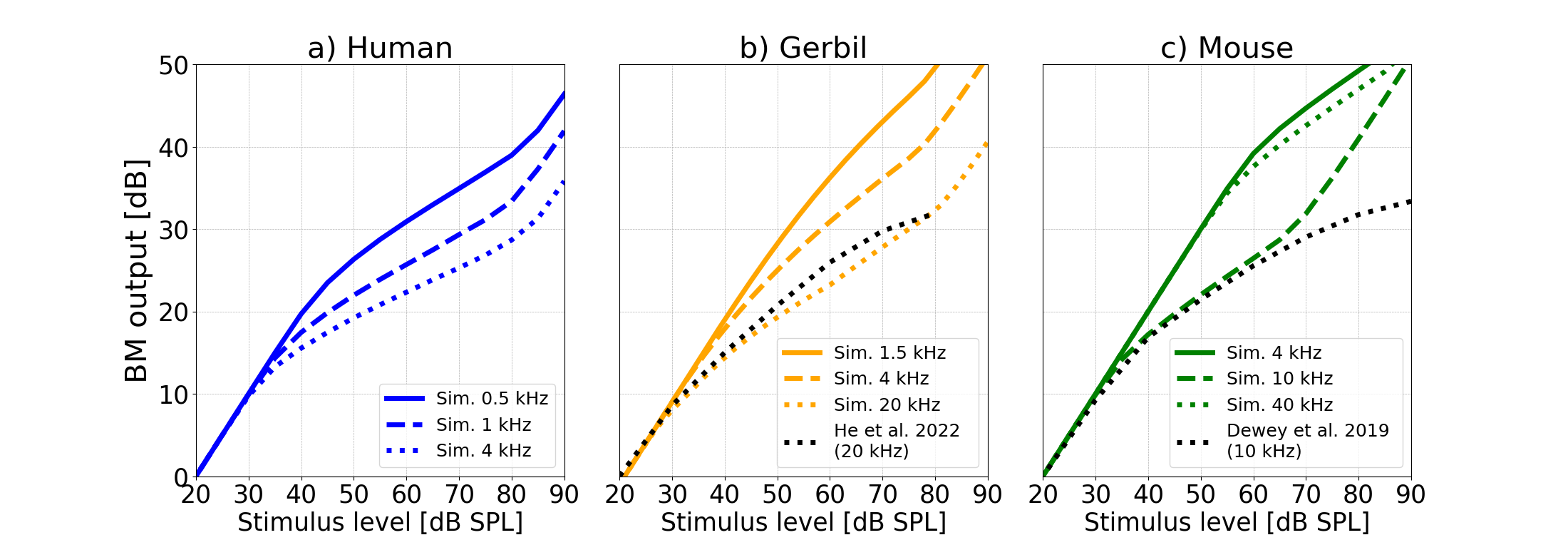}
  \caption{ \textbf{BM velocity growth curves} Model and experimental basilar membrane (BM) responses: BM output velocity as a function of input intensity at the BM location corresponding to each characteristic frequency (CF), illustrating compressive growth. Pure-tone stimuli were presented at the following CFs: (a) 0.5, 1, and 4 kHz for human; (b) 4, 20, and 40 kHz for gerbil (experimental data from \citep{He2022}); and (c) 1.5, 10, and 50 kHz for mouse (experimental data from \cite{Dewey2019}). BM output intensity was normalized to 0~dB at an input level of 20~dB~SPL to facilitate comparison across curves.}
  \label{fig:BM_output}.
\end{figure}

Figure~\ref{fig:BM_output} illustrates the level-dependent BM responses produced by our computational models. We compare BM responses to pure-tone stimulation across models and benchmark them against experimental OCT measurements obtained from gerbil and mouse BM velocity data.

For the gerbil model, figure \ref{fig:BM_output} b shows a close fit between the stimulus level and the BM output at 20 kHz. This fit is also satisfactory in the mouse model at 10 kHz up to approximately 70 dB SPL. At higher stimulus levels, the model and measurements begin to diverge. This discrepancy is caused by the lack of compression in the 1-D TL model and is discussed more in detail in section \ref{BM responses and AN tuning}. In addition, at CFs near the extremes of the hearing range, the animal models tend to exhibit more linear input output functions and therefore reduced compression.

\subsection{AN tuning curves}
\begin{figure}[h!]
  \centering
  \includegraphics[scale=0.3]{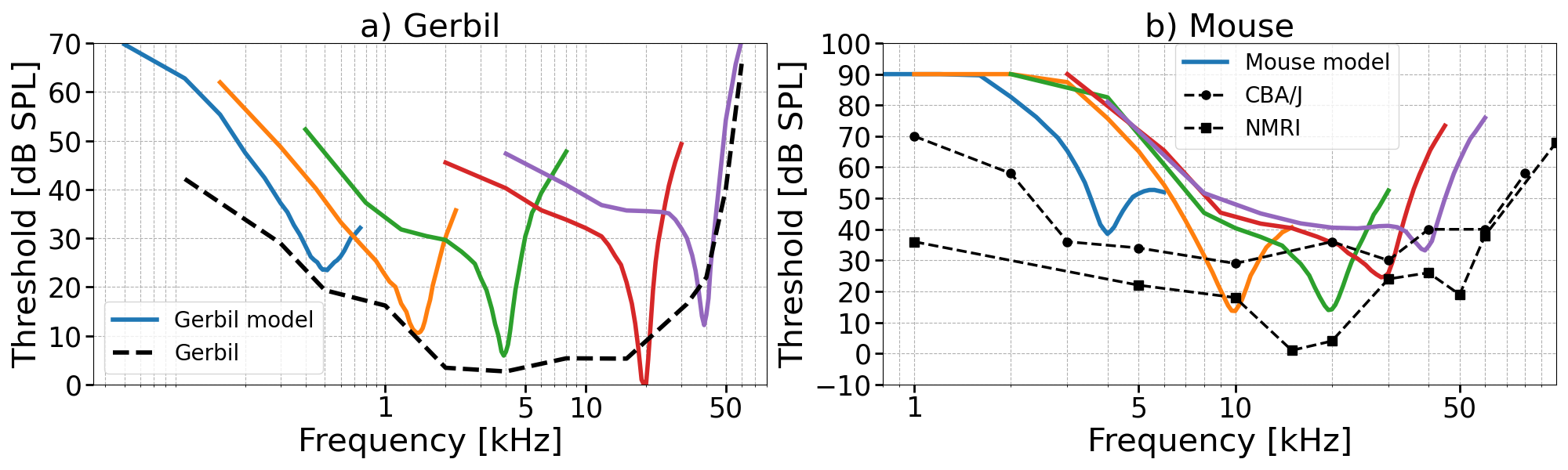}
  \caption{ \textbf{Tuning curves and thresholds} Tuning curves based on high–spontaneous-rate (HSR) auditory nerve fibres in the gerbil (a) and mouse (b) models, shown together with experimental data. Gerbil model thresholds are compared to behavioral thresholds from \cite{Ryan1976}, and mouse model thresholds are compared to behavioral thresholds from \cite{Birch1968,Ehret1974} for the CBA/J and NMRI mouse strains, respectively.}
  \label{fig:Tuning_model}.
\end{figure}

Figure~\ref{fig:Tuning_model} shows the HSR ANF tuning curves for both computational
animal models, along with threshold data reported in the literature. In the gerbil
model, the ANF thresholds align closely with behavioral thresholds reported in 
experimental studies \citep{Ryan1976}. The mouse model exhibits a similar overall 
trend, although simulated ANF thresholds for low frequencies are higher than those reported for the CBA/J and NMRI mouse strains in \cite{Taberner2005}. It is important to note that both ANF and 
behavioral thresholds vary significantly across mouse strains as is visible in Figure~\ref{fig:Tuning_model}.

\begin{figure}[h!]
  \centering
  \includegraphics[scale=0.3]{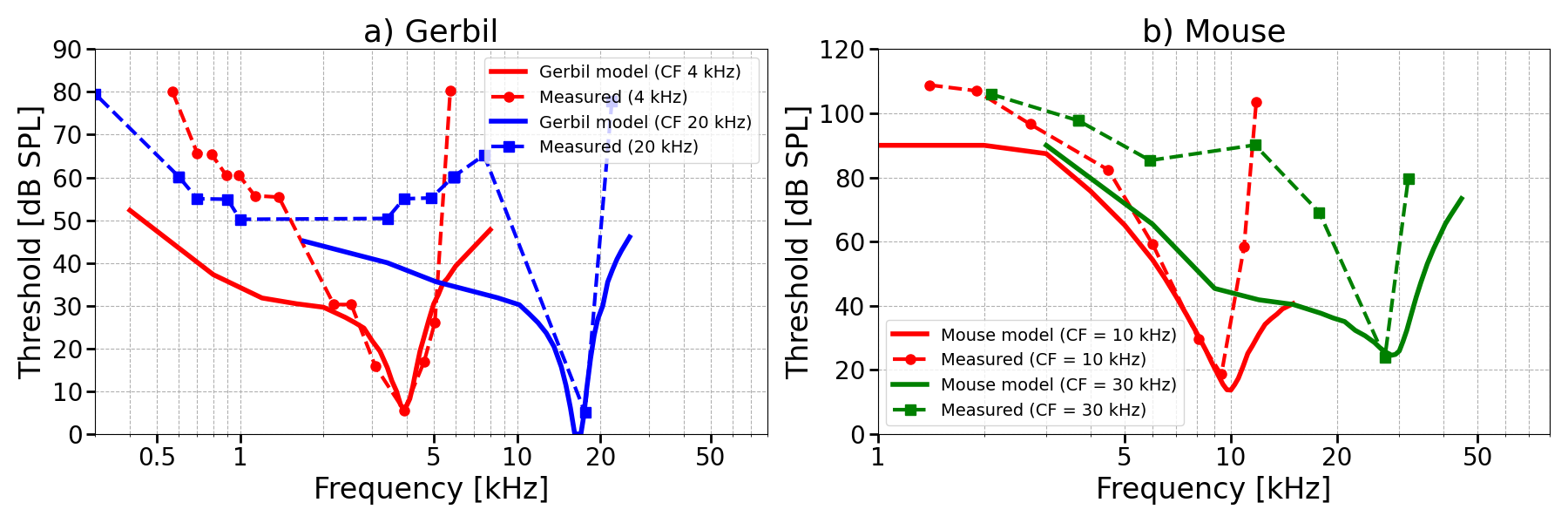}
  \caption{\textbf{Measured and simulated tuning curves} Simulated and measured auditory nerve fibre (ANF) tuning in the a) gerbil and b) mouse for a high–spontaneous-rate fiber with a characteristic frequency of approximately 4 and 17~kHz for gerbil and 10 and 30 kHz for mice. Gerbil measurements from \citep{Daniil,Taberner2005}}.
  \label{fig:Tuning_single}.
\end{figure}

Figure~\ref{fig:Tuning_single} compares measured gerbil and mouse ANF tuning curves with its simulated counterpart. At low stimulus levels, the simulated tuning agrees well with the measurements around the CF. However, the simulated thresholds exhibit a less steep increase with frequency compared to the measured tuning curves of the ANFs. Additionally, the simulated tuning curves demonstrate an underestimated tip-to-tail ratio. These discrepancies increase for CFs located farther from the central BM section where the nonlinearities were calibrated, namely around 4 kHz in the gerbil model and 10 kHz in the mouse model. For instance, the simulated mouse tuning curve at a CF of 30 kHz, with a threshold of approximately 25 dB SPL, is noticeably too broad and its tail too low.

\newpage

\subsection{DPOAEs}
DPOAEs for the three mouse populations are available for comparison and individualization of the computational mouse model. However, DP-levels produced by our TL models are not perfectly within the same range as in vivo measured DPOAEs. For the DP-levels, the measured values ranged from $-25.01$ to $1.85$ dB SPL at $f_2 = 5$ kHz and from $-12.20$ to $35.38$ dB SPL at $f_2 = 45$ kHz. The corresponding simulated values ranged from $-22.87$ to $4.54$ dB SPL at $f_2 = 5$ kHz and from $-18.17$ to $14.81$ dB SPL at $f_2 = 45$ kHz. DPOAEs simulated by the computational mouse model can be compared with the measured DPOAEs by applying the following rescaling:

\begin{equation}
\mathrm{sDP}_\mathrm{{rescaled}} = \frac{\mathrm{sDP} - \mathrm{sDP}_{\mathrm{min}}}{\mathrm{sDP}_\mathrm{{max}} - \mathrm{sDP}_\mathrm{{min}}} (\mathrm{mDP}_\mathrm{{max}} - \mathrm{mDP}_\mathrm{{min}}) + \mathrm{mDP}_\mathrm{{min}}
\label{eq:rescale}
\end{equation}

with:
\begin{itemize}
    \item $\mathrm{sDP}$ = simulated DP-level
    \item  $\mathrm{mDP}$ = measured DP-level
    \item  $\mathrm{sDP}_\mathrm{{max}}$ = simulated DP-value for a flat profile of pole = 0.033
    \item  $\mathrm{sDP}_\mathrm{{min}}$ = simulated DP-value for a flat profile of pole = 0.350
\end{itemize}

The minimal and maximal simulated DPOAEs were assumed to be those produced by the flat starting pole profiles ($\alpha_A$) of 0.350 and 0.033, respectively, as these span the range of poles found for a normal-hearing mouse profile.

\begin{figure}[h!]
  \centering
  \includegraphics[scale=0.2]{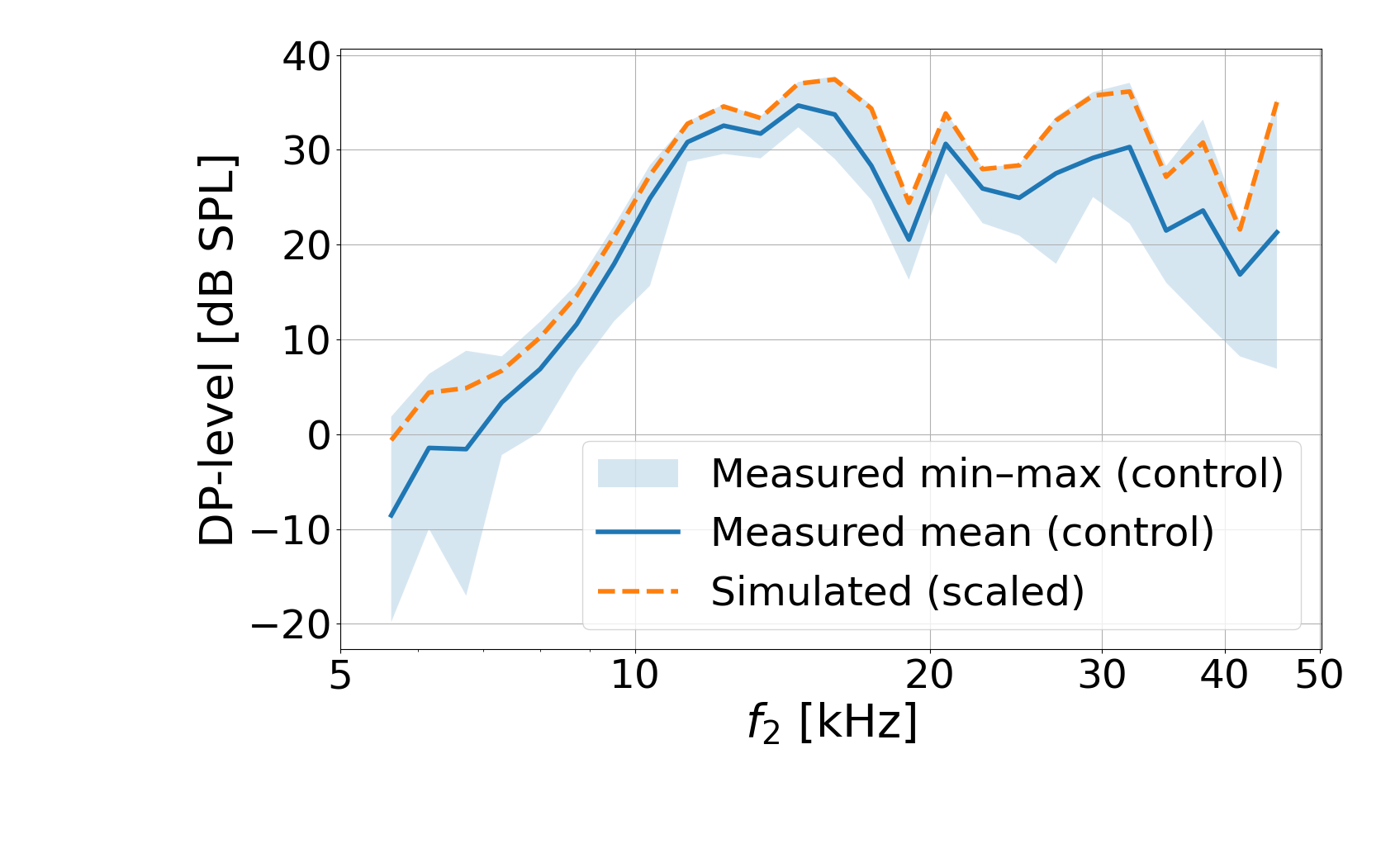}
  \caption{\textbf{Simulated mouse NH DPOAEs} Simulated DPOAE levels for normal-hearing mice are compared with the average measured DPOAE levels, as well as the minimum and maximum values from the control group. DPOAEs were measured using stimulus levels of 55 dB SPL for $f_1$ and 65 dB SPL for $f_2$, with a fixed frequency ratio of $f_2/f_1 = 1.2$. The DP-level is defined as the sound pressure level (dB SPL) measured at the frequency $2f_1 - f_2$. The simulated DPOAEs were linearly rescaled to fall within the measured range using (formula \ref{eq:rescale}).}
  \label{fig:DPOAEs}.
\end{figure}

\begin{figure}[h!]
  \centering
  \includegraphics[scale=0.4]{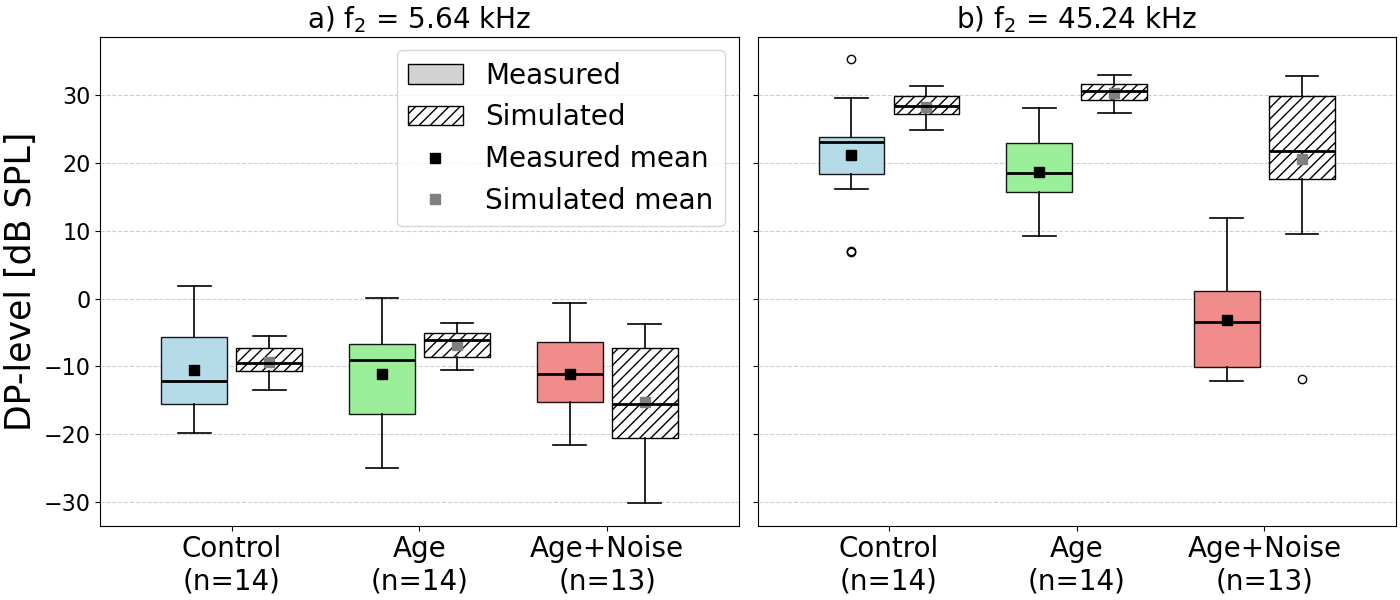}
  \caption{ \textbf{DPOAEs simulations} Simulated and measured mouse DPOAE levels for the control, aged, and noise exposed + aged groups. The simulated DPOAEs were rescaled linearly to fit within the measured range of DPOAEs using this formula ~\ref{eq:rescale}. a) $f_{2}=5.64$ kHz b)  $f_{2}=45.24$ kHz, the latter is also the frequency for which intergroup differences are largest. DPOAEs were measured using stimulus levels of 55 dB SPL for $f_1$ and 65 dB SPL for $f_2$, with a fixed frequency ratio of $f_2/f_1 = 1.2$. The DP-level is defined as the sound pressure level (dB SPL) measured at the frequency $2f_1 - f_2$.}
  \label{fig:DPOAEs_combined}.
\end{figure}

Figure \ref{fig:DPOAEs} shows how the rescaled simulated NH DPOAEs relate to the measured DPOAEs across the mouse cohort. Figures \ref{fig:DPOAEs_combined} a and b compare measured and rescaled simulated DPOAEs for $f_2 = 5.64$~kHz and $f_2 = 45.24$~kHz, respectively, representing the lowest and highest $f_2$ values in the dataset. Simulations were generated after individualizing the model poles based on full-bandwidth DPOAEs. Generally, the simulations match the experimental DPOAEs better at low frequencies. The discrepancy between simulations and experiments becomes larger for high frequencies. Although individual matching at high frequencies is difficult, intergroup differences are reproduced as the noise exposed group clearly shows lower DP-levels at $f_2$ = 45.24 kHz compared to the other groups.

\newpage
\clearpage
\subsection{Auditory Brainstem responses (ABRs)}
\begin{figure}[h!]
  \centering
  \includegraphics[scale=0.3]{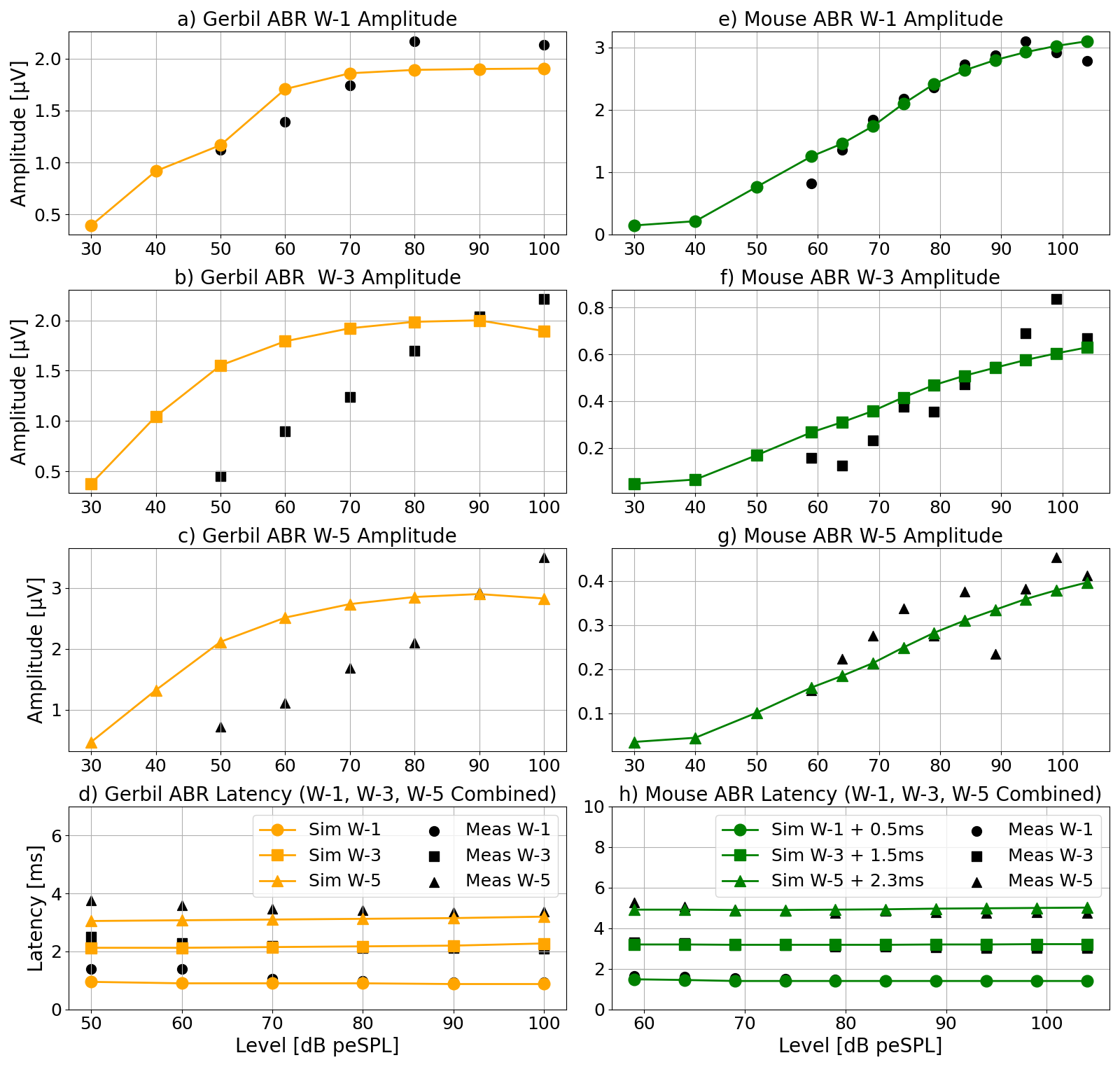}
  \caption{\textbf{ABR simulations} Comparison of simulated and measured click auditory brainstem response (ABR) amplitudes and latencies for gerbil (left column) and mouse (right column). Panels (a)--(c) show the Wave~1 (W1), Wave~3 (W3), and Wave~5 (W5) amplitude--level functions for gerbil, comparing the model output (orange) with published peak-to-peak measurements for condensation clicks (black markers; \citep{burkard1989stimulus}). Panel (d) shows the corresponding latency--level functions for gerbil, with simulated responses (orange) and measured data for clicks (black markers) \citep{burkard1989stimulus}. Panels (e)--(g) present mouse W1, W3, and W5 amplitude--level functions, comparing simulated latencies (green) with reported experimental values (black). Panel (h) displays the mouse latency--level functions for W1, W3, and W5, with simulated latencies (green) plotted against experimental means derived from  data in \citep{Bradmouse2025}.}

  \label{fig:abr_allinone}.
\end{figure}

\begin{figure}[h!]
  \centering
  \includegraphics[scale=0.4]{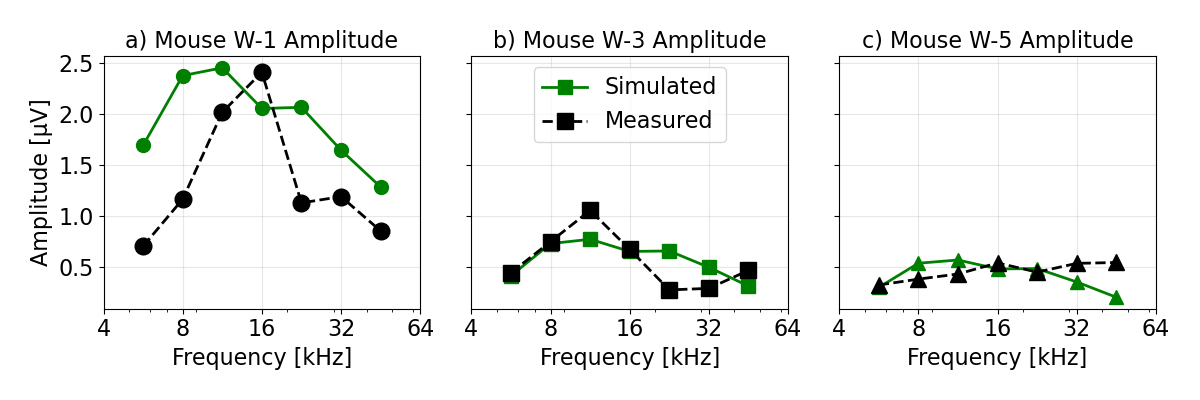}
  \caption{\textbf{ABR as a function of frequency} Simulated and measured ABR peak amplitudes (70 dB SPL) for waves W-1 (a), W-3 (b), and W-5 (c) as a function of stimulus frequency (4--64~kHz). Green symbols and solid lines represent simulated amplitudes, whereas black symbols and dashed lines represent measured amplitudes. Each subpanel corresponds to one ABR wave and uses a consistent marker shape across figures (W-1: circle, W-3: square, W-5: triangle).}

  \label{fig:ABR_mouse_ifo_freq}.
\end{figure}

\begin{figure}[h!]
  \centering
  \includegraphics[scale=0.2]{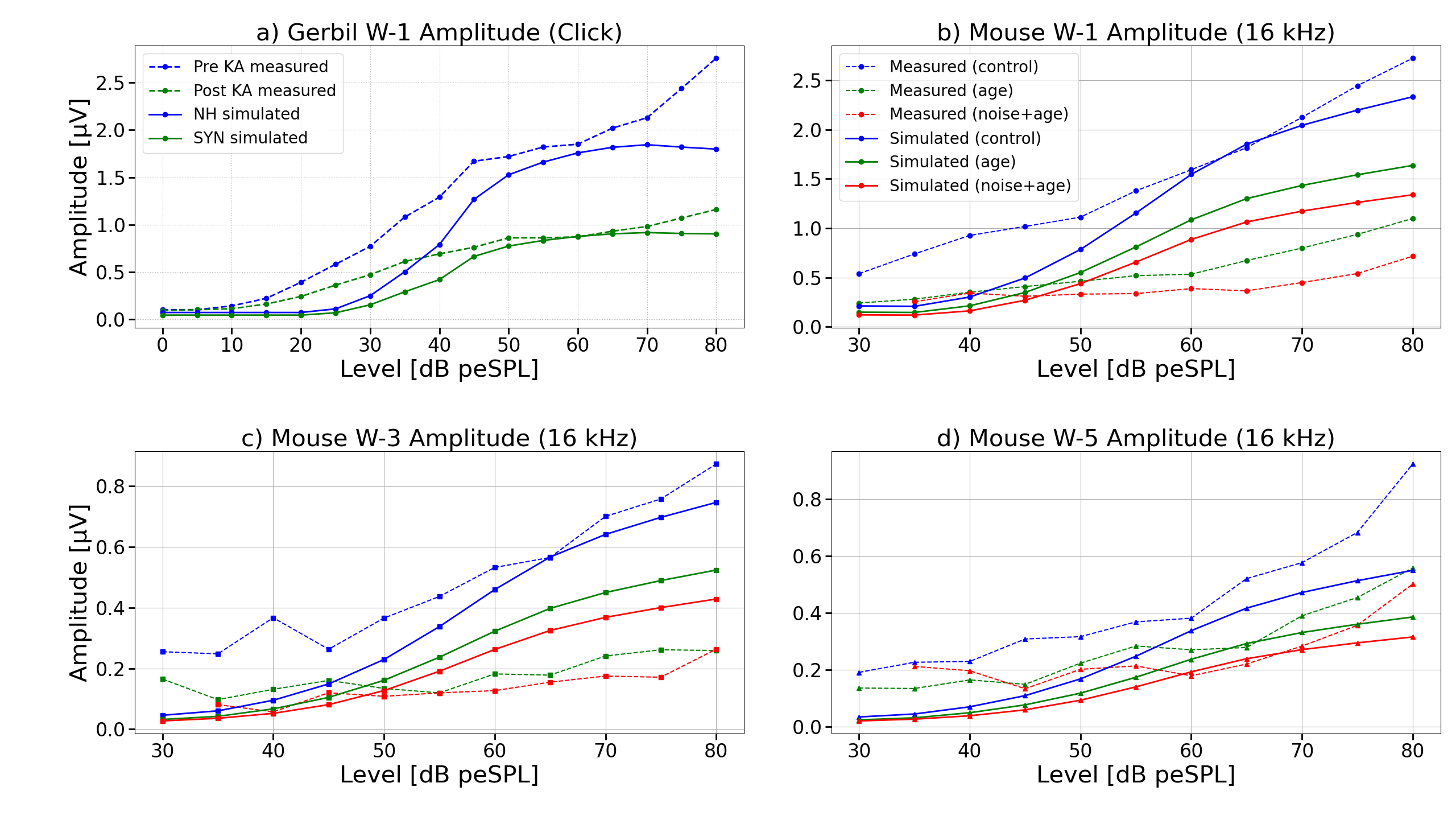}
  \caption{ \textbf{Individualized ABRs} Comparison of simulated and measured auditory brainstem response (ABR) amplitudes gerbil and mouse. Panels (a) shows the Wave~1 (W1) amplitude--level functions for normal hearing (NH), and synaptopathic (SYN) gerbil click ABRs, comparing the model output (full lines) with published (intermittent lines) peak-to-peak measurements for condensation clicks in gerbils before and after kainic acid (KA) treatment. SYN gerbil simulations were done for two thirds of the high spontaneous rate fibers intact and no remaining medium-, or low spontaneous rate fibers. Panels (b)--(d) show the corresponding 16 kHz toneburst ABR amplitude--level functions for mouse, with simulated responses (full lines) and measured data for  (intermitted lines). Means of individualized simulations for the control, age, and age + noise group are plotted against their experimental counterparts derived from Buran et al.\ (2024).}

  \label{fig:indiv_abrs}.
\end{figure}

Simulated click-evoked ABR peak-to-peak amplitudes and latencies for W1, W3, and W5 are shown in Figure~\ref{fig:abr_allinone}, alongside the corresponding measured responses across increasing input levels. Panels~(a)--(c) illustrate the gerbil amplitude growth functions. The simulated W1 input–output curves closely reproduce the measured growth characteristics and overall compressive behavior. For W3 and W5, the simulations exhibit stronger compression above 50~dB~peSPL, whereas the measured curves remain more linear, with only a modest reduction in growth at the highest levels.  The gerbil model clearly overestimates nonlinearities in the CN and IC, suggesting a species specific CN and IC model is required. Panels~(e)--(g) show the mouse amplitude growth functions. Across the tested level range, the simulated curves capture the general shape and level dependence of the measurements, although the empirical W1, W3, and W5 amplitudes display more pronounced compression at higher levels compared to the model predictions. Further, it is also remarkable that the gerbil measurements of W3 and W5 show a stronger linear behavior than is observed in the mouse measurements.

Simulated gerbil ABR latencies closely match the measured data (Figure .~\ref{fig:abr_allinone}~d). In the measurements, W1, W3, and W5 latencies decrease slightly with increasing input level. This trend is accurately reproduced for W5 in the model, whereas the simulated W1 and W3 latencies show almost no level dependent decrease or even tend to show a small growth. A similar decreasing latency trend is observed in the mouse data and in the corresponding simulations (Figure~\ref{fig:abr_allinone}~h). However, the absolute latency error is larger in the mouse model than in the gerbil model. This discrepancy increases across waves, with the smallest deviation for W1, a larger deviation for W3, and the largest for W5. Therefore, constant temporal offsets of 0.5 ms, 1.5 ms, and 2.3 ms were applied to the simulated W1, W3, and W5 responses, respectively, to enable comparison of the level-dependent trends.

Figure~\ref{fig:ABR_mouse_ifo_freq} compares simulated tone-burst ABR peak-to-peak amplitudes with interleaved measured amplitudes \citep{interleaved} for W1, W3, and W5 across CFs from 4 to 64~kHz. The largest discrepancies between simulations and measurements are observed for W1. W1 also exhibits larger variations across frequency in the measurements compared to the simulations. This is also true for W3 but less pronounced. Overall, these results indicate that the model captures the qualitative frequency dependence of the ABR, with quantitatively better agreement for W5 and W3 than for W1.

Figure  \ref{fig:indiv_abrs} shows simulated and measured ABR amplitude--level functions  (peak-to-peak) in gerbil and mouse across hearing pathologies. In panel~(a), the simulated W1 amplitudes for the NH and SYN gerbil conditions follow the overall shape of the published pre-KA and post-KA click ABRs, although the pre-KA data exhibit a sharper rise at higher input levels (above 60 dB peSPL) than observed in the simulations. Panels~(b)--(d) show the corresponding mouse 16~kHz conventionally measured toneburst responses. For the age and noise + age group the difference is large for W1 and W3 at the high input levels. For W1, the simulated amplitude of the control group is too small at low input levels. Overall, the simulations reproduce the general upward trends of the ABR amplitude--level functions and amplitude reduction in KA-treated, aging, and noise exposed animal groups.

\subsection{Envelope-following responses (EFRs)}
\begin{figure}[h!]
  \centering
  \includegraphics[scale=0.3]{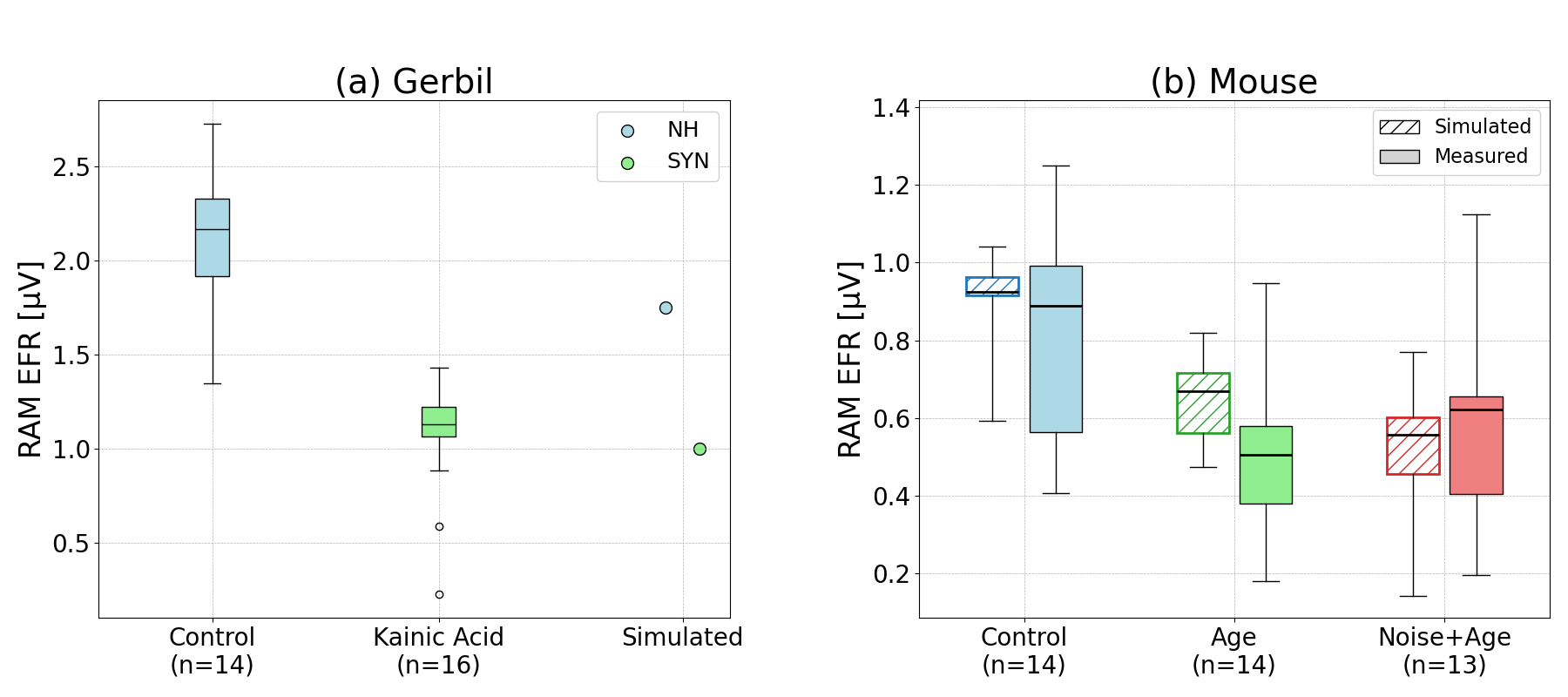}
  \caption{ \textbf{RAM EFRs} a) Measured and simulated gerbil EFR (envelope following response) magnitudes for a RAM (rectangular amplitude modulated) stimulus with a carrier frequency of 4 kHz and a modulation frequency of 116 Hz at 70 dB SPL. For the experimentally measured gerbils there is a control and a kainic acid (KA) treated group, the simulated EFRs contain a normal hearing and a synaptopathic (SYN) group for which no medium-, or low spontaneous rate fibers and all of the high spontaneous rate fibers are present. b) Simulated and measured mouse EFRs to a RAM stimulus with $f_c$ = 8 kHz and $f_m$ = 110 Hz at 70 dB SPL. The mice models were individualized using the poles based on the individual measured DPOAEs and the synapse counts per inner hair cell corresponding to the 8 kHz cochlear frequency region.}.
  \label{fig:EFR_Gerbil_Mouse_with_Measured}.
\end{figure}

RAM-EFRs were simulated using both the gerbil and mouse models. For the gerbil model, simulations were performed for a NH condition with all ANFs intact, as well as for a synaptopathic condition characterized by complete loss of MSR and LSR fibers and no reduction in HSR fibers. Figure  \ref{fig:EFR_Gerbil_Mouse_with_Measured} a) compares the boxplots of the gerbil RAM EFR measurements of the control and KA group to both the NH and SYN (synaptopathy) model simulations. Mouse simulations were performed with models that were fully individualized based on both the DPOAEs and the number of synapses per IHC count  (Figure \ref{fig:ribbons_dpoae_groups}). Boxplots for both the simulated and reference EFR responses can be seen in Figure \ref{fig:EFR_Gerbil_Mouse_with_Measured} b). No rescaling was applied to any of the model RAM EFR simulations. The mouse model generally seems to match RAM EFR measurements. Further, it broadly captures intergroup differences as the age and noise + age group produce lower RAM EFR outputs compared to the control group.

\section{Discussion}
\subsection{Translating model parameters}
The 1-D TL cochlear model proposed by Verhulst et al. (2018) exploits the principle of scaling symmetry, which enables straightforward parameter rescaling and facilitates its translation to non-human species such as the laboratory animals mouse and gerbil, widely used in auditory research. The extensive descriptions of the auditory periphery in gerbils and mice facilitate the construction, calibration, and verification of  these computational models. Nevertheless, it is important to recognize the influence of genetic variability within species. In mice, for instance, different strains can exhibit markedly distinct auditory characteristics  \citep{Taberner2005}. Consequently, it is difficult to define a single representative standard within a species, as is often implied in computational modeling. This limitation should be carefully considered when applying or developing such models. 
\\

Further, we focused primarily on cochlear mechanics without adapting synapse, IHC, or ANF parameters, except for the ANF subtype distribution. Altering these parameters to match animal recordings and possibly vary them across CF could be a next step in the animal model development.

\subsection{BM responses and AN tuning}
\label{BM responses and AN tuning}
Another limitation is the inherent coupling between hearing sensitivity and tuning sharpness in 1-D TL models. As a result, for a given ME filter, only a restricted set of frequency-dependent combinations of tuning and ANF thresholds can be achieved. In addition, ANF thresholds are influenced by the $\kappa_{\mathrm{IHC}}$ parameter, which provides a global gain control that can raise or lower thresholds largely independent of frequency. Consequently, improving tuning and threshold accuracy at one frequency often comes at the expense of the accuracy at another frequency. This trade off can be adjusted by jointly tuning the model pole values and the $\kappa_{\mathrm{IHC}}$ parameter. In this work, we prioritized optimization in the frequency regions most relevant to the animal datasets and measurements considered here.
\\

The key limitation in the 1-D TL framework lies in the reachable dynamic range of cochlear mechanical gain in the model, which is maximally about 35 dB gain for the sharpest filters \citep{verhulstcomputational2018}. How much of this gain is reached in the model is determined by the range between the poles $\alpha^*_A$ and $\alpha^*_P$. $\alpha^*_A$ is derived from $Q_{\mathrm{ERB}}$ measurements, and $\alpha^*_P$ is chosen to be the pole value for which saturation occurs. As the tuning of gerbils and mice is less sharp compared to humans (Figure \ref{fig:panelA}), the starting poles ($\alpha_A^*$) are higher. Since very passive poles (large $\alpha_P^*$ values) increase the feedback delay (Equation \ref{eq:2})  too much and cause a violation of the intensity-invariant zero-crossing, the dynamic range is limited. However, even if very high or very passive pole values were chosen for $\alpha^*_P$ and the intensity-invariant zero-crossing constraint is relaxed, the parallel admittance $Y_{p_n}$ will saturate when $\alpha^*$ reaches the saturation pole. This corresponds to $\rho$ approaching zero and $\delta$ approaching a positive constant value \citep{zweigfinding1991}. This results in the lack of compression that can be seen in Figure \ref{fig:BM_output} for both models, although it is more obvious for mice than for gerbils. This is attributable to the mouse tuning at the examined CF being less sharp than the gerbil tuning at the examined CF, resulting in higher starting poles, which, in turn makes it more difficult for the model to reproduce adequate compression. The limited dynamic range also leads to an underestimation of the tip-to-tail ratio in the simulated AN tuning curves (Figure ~\ref{fig:Tuning_single}). Further, this limitation also contributes to the lacking steepness of AN thresholds for increasing frequencies above CF. However, this could also show that the traveling wave is not impeded adequately in the longitudinal direction. This would be related to the $M_{s0}$ parameter in Equation \ref{eq:1}. The discrepancies of the tip-to-tail ratio and lack of increasing thresholds for higher frequencies do not directly imply that the model does not approximate the $Q_\mathrm{ERB}$ values in Figure \ref{fig:panelA}. The $Q_\mathrm{ERB}$ and $Q_\mathrm{10}$ metric are primarily determined by the tuning at the tip and not by the tails. Nonetheless, for CFs corresponding to cochlear sections far from the center of the BM (Figure ~\ref{fig:Tuning_single} b, CF = 30 kHz), the discrepancy at the tip is too large to approximate the predetermined $Q_\mathrm{ERB}$ values.
\\
 
An underestimated tip-to-tail ratio may result in the premature spread of excitation across CF, occurring at input levels lower than expected. This affects population responses, leading to an overestimation of their amplitudes by the model.

\subsection{ANF subtype distributions}
Given that the goal was to develop animal models that are relevant for SNHL research, realistic distributions of ANF subtypes are crucial. The gerbil AN in particular shows a remarkable ANF profile that can affect AEPs \citep{Bourien2014}. In gerbils, the ratio between HSR, MSR, and LSR fibers varies greatly with CF. This property could be useful for studying the role of different ANF subtypes in sound encoding. Considering the selective loss of LSR and MSR fibers that appears to be associated with synaptopathy \citep{Furman2013Noise, Reijntjes2026HearRes}, it is paramount to correctly characterize and quantify the contribution of different ANF subtypes to the ABRs and EFRs in animal models. Experimental data and computational models of animals with diverse ANF profiles could help clarify the contributions of different ANF subtypes to neural encoding, enabling the identification and treatment of their loss in humans.

\subsection{DPOAEs}
OHC damage can be simulated in the models by modifying the starting poles that define the BM impedance with respect to CF. This approach allows the models to be individualized based on measured DP-levels. Figure \ref{fig:DPOAEs_combined} shows the limited ability of the model to reproduce individual DP-levels. Earlier attempts to individualize the human model based on DPOAEs using the same method, showed a reliable performance at low frequencies and a growing discrepancy between simulations and measurements for higher frequencies \citep{keshishzadeh2021dpoae}. Since mice have a higher frequency range of hearing than humans, this limitation has a more pronounced impact on the overall simulated DP-level accuracy in mice compared to humans. The limited ability of the models to reproduce individualized high frequency DP-levels implies that this TL model lacks accurate high frequency cochlear micromechanics or that a different individualization strategy is needed. However, on a group level, the mouse model is clearly able to at least partially capture the effect of noise induced OHC damage based on DPOAEs (Figure \ref{fig:DPOAEs_combined}).

Another potential source of the discrepancy between the mouse DPOAE simulations and DPOAE measurements may arise from the absence of an advanced ear canal and ME model in this framework. The forward and reverse pressure gain functions between the ear canal and the cochlea show multiple peaks \citep{motallebzadehmouse2021}, which can influence DPOAEs independently of cochlear integrity and OHC function.

\subsection{ABR}
The comparison between simulated and measured ABR responses highlights several important aspects of the model's behaviour across species. For gerbil, the click ABR amplitude growth of W1 was reasonably well captured, matching the compression of the empirical data (Figure \ref{fig:abr_allinone} a). The simulations for W3 and W5, however, showed stronger level-dependent compression than observed experimentally, suggesting that the gerbil model overestimates nonlinearities in the CN and IC, respectively (Figure \ref{fig:abr_allinone} b, c). In mice on the other hand, there seems to be slightly more compression in the data than in the model for the W1 amplitude, which is expected due to the lack of compression at high input levels for the BM movement (Figure \ref{fig:BM_output}). The same lack of compression in the simulations is seen for the W3, and W5 amplitude growth function, suggesting that the mouse model underestimates nonlinearities in the CN and IC, respectively. These differences may indicate that additional species-specific or wave-specific gain-control mechanisms are not yet fully represented in the model. However, it is also remarkable that a stronger linear behavior in the gerbil measurements of W3 and W5 can be observed compared to the mouse measurements. This might indicate differences in CN and IC processing between these species or otherwise reveal differences in measurement methods. In either of those cases, the modeling framework does not provide this differentiation as it seems not to be related to mechanical cochlear processing.

Latency comparisons indicated overall good agreement in the gerbil model. In contrast, the mouse simulations exhibited systematic offsets of approximately 0.5, 1.5, and 2.3 ms for W1, W3, and W5, respectively. This suggests that a more detailed CN and IC model, specifically parameterized for the mouse and accounting for these offsets, may be required to accurately reproduce ABR latencies.

In Figure \ref{fig:indiv_abrs} click-evoked ABR W1 simulations in NH gerbils exhibited strong compression above 50~dB peSPL, whereas the measured responses for control animals showed renewed amplitude growth above 70~dB peSPL. It should be noted that evoked potentials at high sound levels are difficult to predict because of large excitation spreads and the recruitment of a very large number of ANFs \citep{Lee2019JNeurophysiol}. The lack of amplitude growth at high levels could be related to the lacking tip-to-tail ratio in the tuning curves. Further, this discrepancy may indicate that the model's LSR fiber threshold is too low, or more generally, that the AN model requires a broader diversity of ANF profiles. 
Also, the shallower rise of the measured amplitude above 70 dB peSPL in gerbils post KA administration compared to pre KA administration measurements enforces the assumption that LSR and MSR fibers are more prone to KA-induced cochlear synaptopathy.

The frequency dependence of mouse toneburst ABR amplitudes were simulated fairly well, although measurements show a larger variability across frequency (Figure \ref{fig:ABR_mouse_ifo_freq}).

\subsection{EFR}
Both the gerbil and the mouse model were capable of simulating RAM EFRs with realistic outputs (Figure \ref{fig:EFR_Gerbil_Mouse_with_Measured}). When all MSR, LSR, and none of the HSR fibers were removed, the simulated gerbil RAM EFRs showed a realistic decline (Figure \ref{fig:EFR_Gerbil_Mouse_with_Measured} a). This choice was made assuming the selective loss of LSR and MSR fibers associated with synaptopathy \citep{Furman2013Noise}. The mouse model produced RAM EFRs with magnitudes well within the range of the measurements, although the simulated output range was smaller than the measured range (Figure \ref{fig:EFR_Gerbil_Mouse_with_Measured} b).  After individualization using DP-levels and synapse counts, the mouse RAM EFR simulations reproduced intergroup differences comparable to those observed in the measurements when comparing the control group with the age groups. This indicates that the mouse model captures the relationship between OHC damage and synapse loss on the one hand, and EFR outcomes on the other.
   
\section{Conclusion}
This paper described how a biophysical model of human auditory processing can be translated to commonly used lab animals while maintaining the concepts of cochlear scaling symmetry and the main model principles. Adjusting the model parameters to capture a higher audible frequency range and modifying the tuning of the basilar membrane introduced a lack of compression due to the initial 1-D TL model structure. Because it is difficult to simultaneously calibrate tuning, ANF thresholds and BM compression across the full frequency range, model calibration prioritized the range of frequencies and stimulus levels most relevant for in vivo measurements.

Mouse model poles can be individualized using measured DPOAEs, and by doing so, it can capture the aspects of OHC damage at a group level. Individual differences on the other hand are not reproduced faithfully. Further, synapse counts are easily applied for model individualization.

By providing species-specific, yet parallel, biophysical models of the auditory periphery, this framework offers a direct way to relate invasive rodent data (e.g., synapse counts and ANF recordings) to non-invasive physiological measurements that are routinely obtained in humans. In doing so, it refines the interpretation of experimental outcomes across species and supports the development of diagnostic measures for cochlear synaptopathy and other forms of SNHL. As these models continue to mature and are integrated with individualized human modelling approaches, they could increasingly support in silico exploration of cochlear pathologies, candidate diagnostics, and interventions, thereby complementing and, where possible, reducing the need for additional animal experiments.

\section{Acknowledgments}
This work is supported by:
ERA-NET project “CoSySpeech”  CoSySpeech (G0H6420N),  FWO Audimod (G068621N), ANR-20-NEUR-0008 and NIH on Deafness and Other Communication Disorders (\#R01DC020423).

\end{document}